\documentclass[10pt, conference, letterpaper]{IEEEtran}
\usepackage{graphicx}
\usepackage{epstopdf}
\DeclareGraphicsExtensions{.eps}
\graphicspath{ {figs/} }
\usepackage{amsmath}
\usepackage{amsfonts}
\usepackage{multirow}
\usepackage{wrapfig}
\usepackage{subcaption}
\usepackage[font=small]{caption}

\usepackage{array}
\newcolumntype{P}[1]{>{\centering\arraybackslash}p{#1}}
\newcolumntype{M}[1]{>{\centering\arraybackslash}m{#1}}
\usepackage{algorithm}
\usepackage{algpseudocode}
\usepackage [english]{babel}
\usepackage [autostyle, english = american]{csquotes}
\MakeOuterQuote{"}

\DeclareSymbolFontAlphabet{\mathrm}    {operators}
\DeclareSymbolFontAlphabet{\mathnormal}{letters}
\DeclareSymbolFontAlphabet{\mathcal}   {symbols}
\DeclareMathAlphabet      {\mathbf}{OT1}{cmr}{bx}{n}
\DeclareMathAlphabet      {\mathsf}{OT1}{cmss}{m}{n}
\DeclareMathAlphabet      {\mathit}{OT1}{cmr}{m}{it}
\DeclareMathAlphabet      {\mathtt}{OT1}{cmtt}{m}{n}
\DeclareSymbolFont{operators}   {OT1}{cmr} {m}{n}
\DeclareSymbolFont{letters}     {OML}{cmm} {m}{it}
\DeclareSymbolFont{symbols}     {OMS}{cmsy}{m}{n}

\usepackage[table]{xcolor}

\usepackage{booktabs}
\usepackage{siunitx}
\usepackage{fancyhdr}
\fancypagestyle{plain}{%
  \fancyhf[C]{Accepted to IEEE INFOCOM 2019}
  \fancyfoot{}
}
\begin{document}
\title{ORACLE: Optimized Radio clAssification through Convolutional neuraL nEtworks }
\author{\IEEEauthorblockN{Kunal Sankhe, Mauro Belgiovine, Fan Zhou, Shamnaz Riyaz,  
Stratis Ioannidis, 
and Kaushik Chowdhury} 
\IEEEauthorblockA{ Electrical and Computer Engineering Department, Northeastern University, Boston, MA, USA}\\
\vspace{-0.7cm}
}

\maketitle
\thispagestyle{plain}

 \begin{abstract}
  
This paper describes the architecture and performance of ORACLE, an approach for detecting a unique radio from a large pool of bit-similar devices (same hardware, protocol, physical address, MAC ID) using only IQ samples at the physical layer. ORACLE trains a convolutional neural network (CNN) that balances computational time and accuracy, showing 99\% classification accuracy for a 16-node USRP X310 SDR testbed and an external database of $>$100 COTS WiFi devices. Our work makes the following contributions: (i) it studies the hardware-centric features within the transmitter chain that causes IQ sample variations; (ii) for an idealized static channel environment, it proposes a CNN architecture requiring only raw IQ samples accessible at the front-end, without channel estimation or prior knowledge of the communication protocol; (iii) for dynamic channels, it demonstrates a principled method of feedback-driven transmitter-side modifications that uses channel estimation at the receiver to increase differentiability for the CNN classifier. The key innovation here is to intentionally introduce controlled imperfections on the transmitter side through software directives, while minimizing the change in bit error rate. Unlike previous work that imposes constant environmental conditions, ORACLE adopts the `train once deploy anywhere' paradigm with near-perfect device classification accuracy.  

 \end{abstract}

\IEEEpeerreviewmaketitle
\vspace{-0.2cm}
\section{Introduction}
\vspace{-0.2cm}
Sensing the wireless spectrum and identifying active radios within the bands of interest directly impacts spectrum usage. This paper takes the first step in distinguishing radios in a shared spectrum environment by using machine learning to detect characteristic reference signatures embedded in their transmitted electromagentic waves, a process known as \textit{RF fingerprinting}. Our goal is to achieve this with information that can be leveraged at the radio hardware front-end. We separately consider situations where the channel is unchanging between training and validation (idealized) and when the channel is dynamic (practical). The key innovation in our approach, termed ORACLE, is that it learns the unique modifications present within the in-phase (I) and quadrature-phase (Q) samples that are introduced in the signal as it passes through the transmitter chain. ORACLE uses Convolutional Neural Networks (CNNs) to learn and then identify individual radios through device-specific variations contributed by the inherent randomness in the manufacturing process. These so called \textit{imperfections} are present within the analog
components (digital-to-analog converters, band-pass filters, frequency
mixers and power amplifiers) that compose a typical transmission chain, differentiating radio devices even if their manufacturer and make/model are identical.  

\begin{figure}
  \centering
  \includegraphics[width=0.8\linewidth]{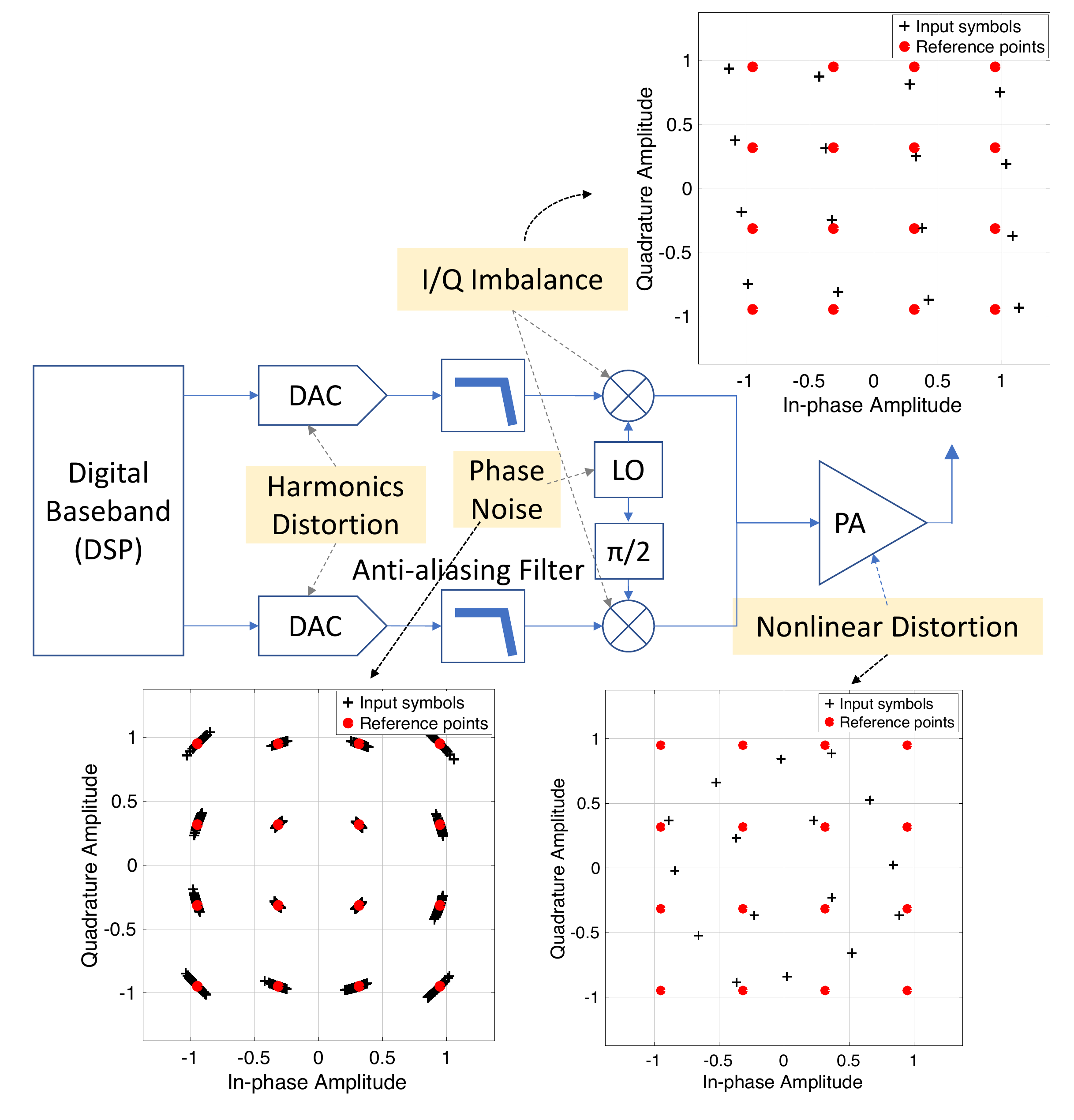}
   \caption{\small \label{fig:arch}Typical transceiver chain with various sources of RF impairments.}\vspace{-0.4cm}
\end{figure}
\subsection{Signatures contained within IQ samples}
Radio fingerprinting involves extracting unique patterns (or features) across the protocol stack that can be used as device signatures. Indeed,  physical (PHY) layer, medium access control (MAC) layer, and upper layers have been  utilized for radio fingerprinting~\cite{fingerprinting}. However, simple unique identifiers such as IP addresses, MAC addresses, international mobile station equipment identity (IMEI) numbers can easily be spoofed. Location-based features such as radio signal strength (RSS), angle of arrival (AoA) and channel state information (CSI) are susceptible to mobility and environmental changes. ORACLE, instead, focuses on those transmitter features that are inherent to a device's hardware makeup, which are unchanging and cannot be easily replicated by malicious agents.

Fig.~\ref{fig:arch} indicates an example scenario of these so called transmitter signatures (rigorously studied in Sec. III) for 16-QAM constellation. The red circles indicate the ideal constellation points formed by the I (x-axis) and Q (y-axis) samples, and the black crosses indicate actual constellation points that are shifted due to a specific type of hardware imperfection. Practical transmitters have a combination of these shifts that form their unique signatures, though we show only three plots caused by  IQ imbalance, nonlinear distortion and phase noise in the figure. ORACLE aims to learn and intentionally modify some of these features on the transmitter through USRP Hardware Driver (UHD) software API commands, thereby enhancing identifiability/classifier efficiency. We note that ORACLE can be easily used in conjunction with other existing and higher layer classification approaches.

\subsection{Machine learning for RF fingerprinting in ORACLE}
Machine learning (ML) techniques have shown great promise in image and speech identification problems, and are steadily gaining traction in applications within the wireless domain. 
ORACLE is solely built on a convolutional neural network architecture that has not only seen success in the above areas, but has also been previously used for modulation~\cite{OSheaC16} and protocol identification~\cite{selim2017spectrum}. ORACLE adopts a stagewise approach towards achieving practical classification. We attain this in the first step by demonstrating 99\% accuracy on an externally obtained data set of 100+ COTS WiFi radios (not all of which are bit-similar), as well as on our testbed of 16 bit-similar USRP X310 radios that we configure to be exactly similar in terms of waveforms generated (same 802.11a PHY frame, modulation/protocol/mac ID).  

\subsection{The ORACLE approach}
For radios operating in a channel-invariant environment (henceforth referred to as a \textit{static} channel), ORACLE identifies radios by using only raw IQ samples. It neither estimates the channel, nor does it use any prior knowledge of the protocol being used. However, its performance degrades 
if the operating environment of the radio is changed. This is because the wireless channel often has a dominant impact on the transformation of the IQ samples in the complex plane. When the channel is varying (henceforth referred to as a \textit{dynamic} channel), ORACLE is trained with complex demodulated symbols instead of raw IQ samples. This approach negates the effect of the channel while retaining the effect of hardware impairments only. Here we make an interesting observation: training with demodulated symbols makes low-end SDRs (such as the Ettus N210 USRP) robust to channel variations. However, high-performance SDRs (such as the X310 USRP) that are manufactured with components with lower variability need an additional step. For such high-end bit-similar devices, ORACLE has a principled method for intentionally introducing impairments to increase differentiability while minimizing the bit error rate (BER) for each transmitter. The key insight here is that controlled addition of impairments in a bit-similar radio generates a unique pattern in the demodulated signal at the receiver, which is independent of channel variations. 

In summary, the main contributions of this paper are:
 
\noindent$\bullet$ We study the different causes of transmitter-side reference signatures, and visualize their impact on the IQ constellation space. We identify specific features that are amenable to fine tuning by the receiver feedback using software APIs.

\noindent$\bullet$  Using an SDR testbed and external database of 100+ devices, we propose the design of ORACLE, which includes a robust CNN architecture returning $>$99\% device classification accuracy on static channels using only raw 1/Q samples.

\noindent$\bullet$ We propose and implement an enhanced design of ORACLE on USRP X310 radios, that systematically introduces controlled impairments to increase differentiability in high-end bit-similar SDRs, while ensuring the added BER at a common receiver is minimized. This is a critical step towards `train once deploy anywhere' paradigm that allows robust CNN learning under realistic channel variations.


\section{Related work}

\begin{table}[!t]
\setlength{\extrarowheight}{1pt}
\label{table:history}
\caption{Machine learning approaches for device fingerprinting.}
\begin{centering}
 \begin{tabular}{|| >{\centering\arraybackslash}m{2.7cm} | >{\centering\arraybackslash}m{5.0cm} ||}
 \hline
  Publication & Approach  \\ 
  \hline 
Franklin \textit{et al} \cite{franklin}  &  Master DB of signatures for wireless device driver fingerprinting  \\\hline
Gao \textit{et al} \cite{Gao} &  Master DB of signatures for AP fingerprinting \\\hline
Kennedy \textit{et al} \cite{algo_class3}  &  k-NN based transmitter fingerprinting \\\hline
Brik \textit{et al} \cite{algo_class2} &  SVM based NIC identification \\\hline
Radhakrishnan \textit{et al} \cite{algo_class5} &  ANN based wireless device identification \\\hline
O'Shea, \textit{et al} \cite{OSheaC16}  &  CNN based Modulation recognition  \\\hline
Chen, \textit{et al} \cite{chenpassive} &  Infinite Hidden Markov Random Field based classification \\\hline
Nyugen, \textit{et al} \cite{unsuper7}  &  Infinite Gaussian Mixture Model based device classification \\\hline
 \hline
\end{tabular}
\end{centering}
\vspace{-0.3cm}
\end{table}

While there exists a vast literature on the theory and applications of ML, we only review works that are directly relevant to the problem of RF fingerprinting, and within it, mainly supervised learning. Unsupervised learning, on the other hand, is  effective when there is no prior label information about devices. For e.g., in \cite{chenpassive}, an infinite Hidden Markov Random field (iHMRF)-based online classification algorithm is proposed for wireless fingerprinting using unsupervised clustering techniques and batch updates. Transmitter characteristics are used in \cite{unsuper7} where a non-parametric Bayesian approach (namely, an infinite Gaussian Mixture Model) classifies multiple devices in an unsupervised, passive manner. However, in our approach we generate real data for each device independently; hence, labeling the device specific dataset is an inexpensive task. Given the ground truth to facilitate model creations, we follow the supervised learning paradigm, where a large collection of labeled samples are applied for training, prior to network deployment. There are two main approaches in this form of learning:
\vspace{-0.1cm}
\subsection{Similarity-based}
Similarity measurements involve comparing the observed signature of the given device with the references present in a master database. In \cite{franklin}, a passive fingerprinting technique is proposed that identifies the wireless device driver running on an IEEE 802.11 compliant node by collecting traces of probe request frames from the devices. A supervised Bayesian approach is used to analyze the collected traces and generate the device driver fingerprint. Gao et al.~\cite{Gao} describe a passive blackbox technique, that uses TCP or UDP packet inter-arrival time to determine the type of access points using wavelet analysis. However these techniques rely on prior knowledge of vendor specific features. 
\vspace{-0.1cm}
\subsection{Classification-based}
\subsubsection{Conventional}  This form of classification examines a match with pre-selected features using domain knowledge of the system, i.e., the dominant feature(s) must be known a priori.
Kennedy et al.~\cite{algo_class3} propose classification by extracting the known preamble within a packet and
computing spectral components. A set of log-spectral-energy features are given as input to the k-nearest neighbors (k-NN) discriminatory classifier. PARADIS \cite{algo_class2} fingerprints 802.11 devices based on modulation-specific errors in the frame using SVM and k-NN algorithms
with an accuracy of 99\%. In \cite{algo_class5}, a technique for physical device and device-type classification called GTID using artificial neural networks is proposed that exploits variations in clock skews as well as hardware compositions of the devices. However, as multiple different features  are used, selecting the right set of features is a challenge. This also causes scalability problems when large number of devices are present, leading to increased computational complexity in training.
\subsubsection{Deep Learning}
Deep learning offers a powerful framework for learning complex functions, leverages large datasets, and greatly increases the the number of layers, in addition to neurons within a layer.
O'Shea and Corgan \cite{OSheaC16} and O'Shea and Hoydis \cite{deep-tim} apply deep learning at the physical layer, specifically focusing on modulation recognition using IQ samples and convolutional neural networks. They classify 11 different modulation schemes. However, this approach does not identify a device like ORACLE, but only the modulation type used by the transmitter.

 
To the best of our knowledge, ORACLE is the first work that allows training a CNN for bit-similar device identification  such that the same classifier may operate in unknown/dynamic channel conditions without the need for new trials. 


\vspace{-0.1cm}

\section{A closer look at device signatures}
In this section, we first study RF hardware impairments that cause variations in IQ samples, resulting in a unique \textit{signature} for each device. We focus on IQ imbalance and DC offset, the two impairments  that (i) are independent of the environment, and (ii) do not apply only in context of a specific transmitter-receiver pair (as opposed to, say, relative phase offset). Then, we present a method of introducing controlled impairments using GNU Radio UHD API at the receiver. Subsequently, we explain the  experimental testbed setup for trace data collection. 

\subsection{RF impairments}
\label{sec:impairments}
Using the MATLAB Communications System Toolbox, we simulate a typical wireless communications processing chain (see Fig.~\ref{fig:arch}, with the shifts in the received complex valued IQ samples), and then modify the ideal operational blocks to introduce RF impairments, typically seen in actual hardware implementations. This allows us to individually study the IQ imbalance, DC offset, phase noise, carrier frequency offset and nonlinear distortions of power amplifier. In this paper, we focus on the two impairments (IQ imbalance and DC offset) owing to space constraints, though our approach can be trivially extended for others as well. 

\noindent $\bullet$\textbf{IQ imbalance:} 
Quadrature mixers are often impaired by gain and phase mismatches between the parallel sections of the RF chain dealing with the I and Q signal paths. The mismatch in their gains causes amplitude imbalance, whereas phase deviation from $90^{\circ}$ in the quadrature signal results in phase imbalance. IQ imbalance varies only with frequency due to frequency-dependent low pass filters, and thus, it carries a unique signature of a transmitter for that frequency.   
\\
\noindent $\bullet$\textbf{DC offset:} 
This is caused within the quadrature mixers due to the finite isolation between Local Oscillator (LO) and RF ports of a mixer, and a direct feedthrough from the LO signal often gets coupled to the output. \\

\subsection{Software-based control of impairments}
We first explain the use of self-calibrations utilities provided by Ettus to set IQ imbalance and DC offset in the transmitter chain using GNU Radio functions. 

\label{subsubsec:utilities}
 $\newline$
$\noindent$ $\bullet$\ \textbf{$\textit{IQ imbalance compensation}$:}
Let $s(t)\in \mathbb{C}$ be the transmitted baseband complex signal at time $t$ before being distorted by IQ imbalance. Then, the distorted baseband signal in the time domain is:
\begin{equation}
s_d(t)= \mu_t s(t) + v_t s^*(t),
\vspace{-0.1cm}
\end{equation}
where the distortion parameters $\mu_t$ and $v_t$ are related to amplitude and phase imbalances in the I and Q paths of the quadrature mixer in the transmitter chain. 

The simplified model of these distortions parameters can be written as $\mu_t=\cos{\big(\theta_t/2\big)}+j\alpha_t 
\sin{(\theta_t/2\big)}$ and $v_t=\alpha_t\cos{\big(\theta_t/2\big)} - j 
\sin{(\theta_t/2\big)}$,
where $\alpha_t$ and $\theta_t$ are the amplitude and phase imbalance between the I and Q signal paths at the transmitter, respectively. The phase imbalance is any phase deviation from the ideal $90^{\circ}$. The amplitude imbalance is defined as $\frac{\alpha_I- \alpha_Q}{\alpha_I + \alpha_Q}$, where $\alpha_I$ and $\alpha_Q$ are the respective gain amplitudes on the I and Q paths.

\begin{figure}
  \centering
  \includegraphics[width=0.8\linewidth]{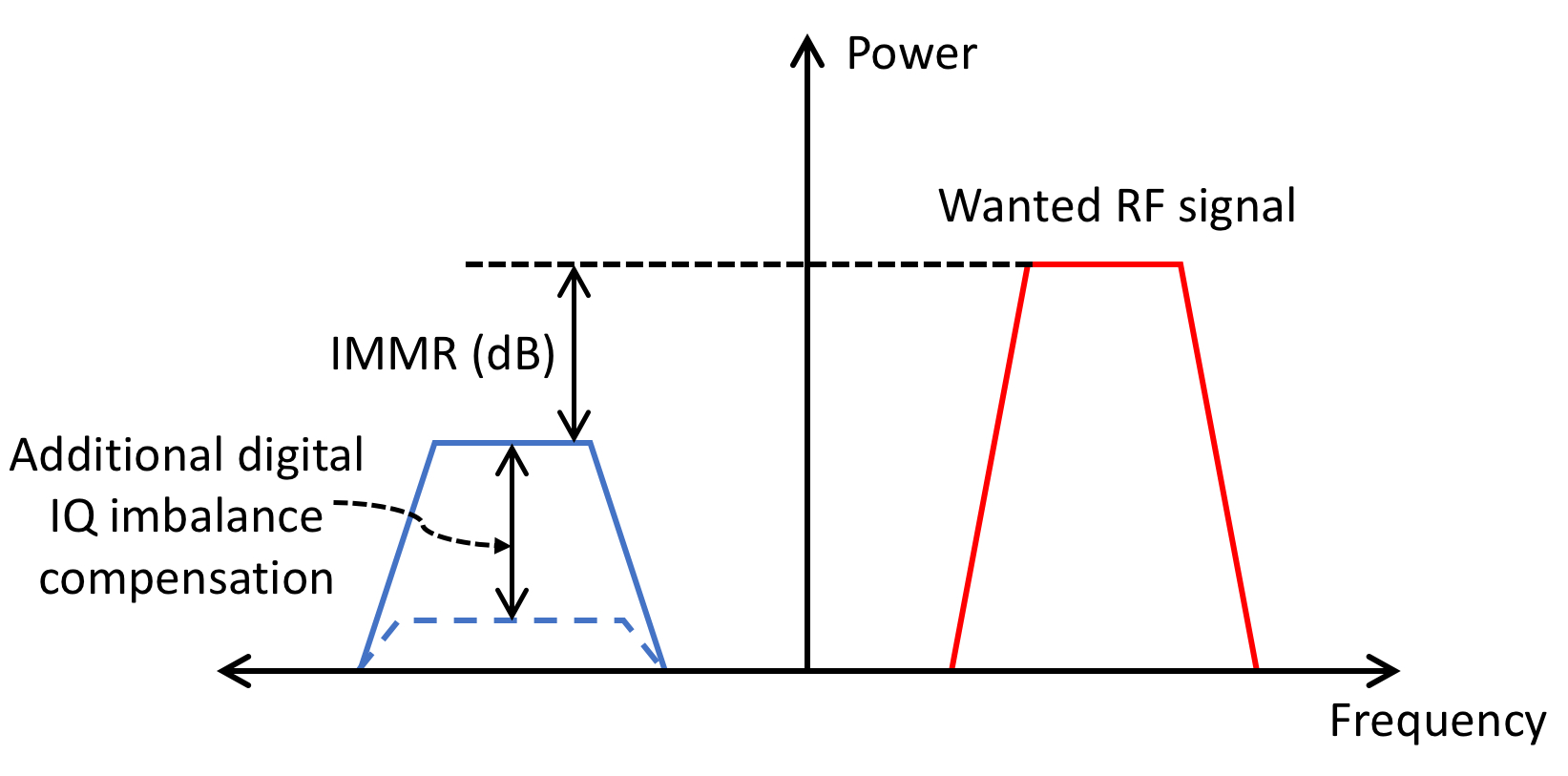}
   \caption{\small \label{fig:IMMR} Effect of IQ imbalance quantified through IMRR.}\vspace{-0.4cm}
\end{figure}

IQ imbalance causes interference in the signal by generating its image at a mirror frequency. It is quantified by measuring the power of the image with respect to the desired signal, also called as Image Rejection Ratio (IMRR), as shown in Fig.~\ref{fig:IMMR}. The IMRR is calculated by sending a complex sinusoidal $e^{jwt}$, and by taking ratio of the power of the signal at the image frequency $(-w)$ and desired frequency $(w)$. Thus, IMMR for amplitude imbalance $\alpha_t$ and phase difference of $\theta_t$, is given by:
\begin{equation}
\textit{IMRR}=\frac{\gamma_t^2 +1 -2 \gamma_t \cos{\theta_t}}  {\gamma_t^2 + 1 + 2 \gamma_t \cos{\theta_t}},
\vspace{-0.1cm}
\end{equation}
where $\gamma_t=\alpha_t+1$.

While many theoretical time and frequency domain methods allow compensation for the IQ imbalance, we use the Ettus provided UHD calibration utility $\texttt{uhd\_cal\_tx\_iq\_balance}$. It performs a calibration sweep over a range of frequencies checking the transmission path signal leakage into the receive path. 

At runtime, the UHD software automatically applies the correction, typically a single complex factor, to the transmit chain of the RF daughterboard. \texttt{SampIn} is the complex sample input to the block, \texttt{SampOut} is the complex sample output and \texttt{Corr} is the correction factor. 
For a given value of correction factor, a single frequency tone is transmitted, and the power of the desired tone and the image tone are measured to compute IMMR. 

We modified this utility to record the correction factors and the corresponding IMMR. Table \ref{Tabel:IQBalance} shows a snapshot of the recorded IMMR levels for USRP X310 radio at a center frequency of 2.45 GHz. 
 
 \begin{figure}
   \includegraphics[width=\linewidth]{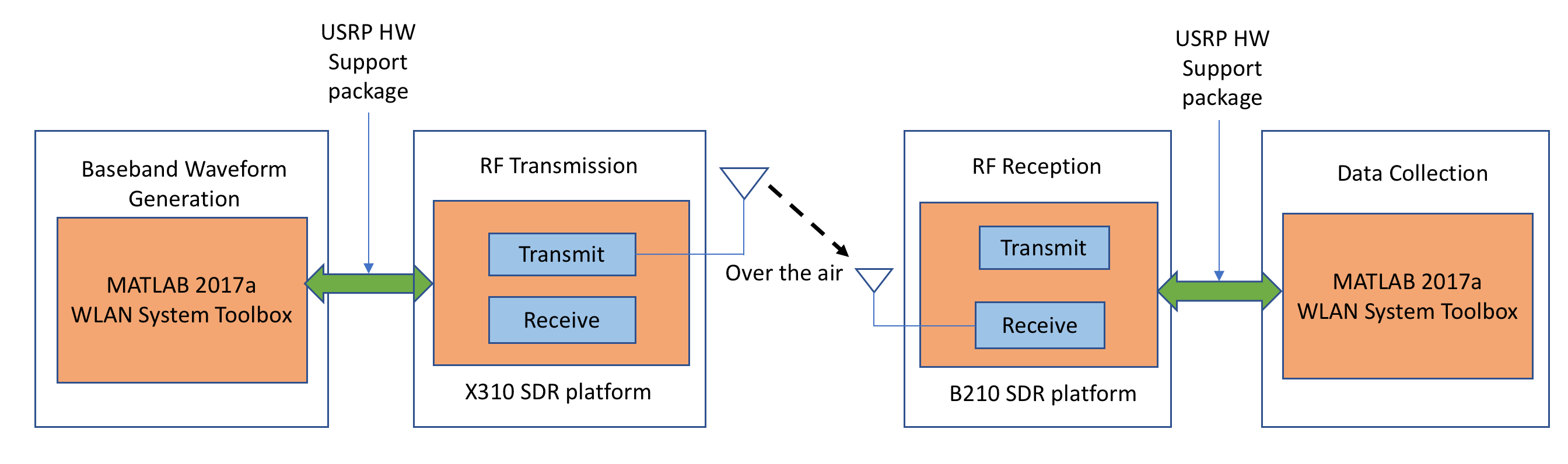}
   \caption{Experimental setup for data collection using SDR. }
   \label{fig:dataCollect} \vspace{-0.2cm}
\end{figure}

\begin{table}[!t]
	\caption{A snapshot of IMMR levels of IQ imbalance recorded using $\texttt{uhd\_cal\_tx\_iq\_balance}$ utility }
	\label{Tabel:IQBalance}
\centering
\small
\begin{tabular}{||  >{\centering\arraybackslash}m{1.2cm} | >{\centering\arraybackslash}m{1.3cm} | >{\centering\arraybackslash}m{1.4cm} | >
{\centering\arraybackslash}m{1.4cm} | >{\centering\arraybackslash}m{1.2cm} ||}
\hline
 Correction real & Correction imag. & Power of main tone & Power of image tone & IMMR (dB) \\
\hline \hline
 $-0.272$ & $-0.636$& $-49.036$ & $-66.138$ & $-17.102$\\ \hline
  $-0.636$ & $-0.636$ & $-48.852$ &$-66.306$ & $-17.454$\\ \hline
 $-0.454$ & $-0.0909$ & $-49.091$  & $-67.326$ & $-18.235$\\ 
\hline
\end{tabular}
\vspace{-0.4cm}
\end{table}

$\noindent$ $\bullet$\ \textbf{$\textit{DC offset compensation}$:}
DC offset results in a large spike in the center of the spectrum. By measuring the power of the main tone at the DC frequency, we can measure the amount of DC offset. A UHD calibration utility $\texttt{uhd\_cal\_tx\_dc\_offset}$ uses a single complex factor to correct DC offset level. It finds the best correction factor that minimizes the power of the DC tone. Again, by modifying the utility, we record the levels of DC offset level for the correction factor. 

We use the open-source GNU Radio companion (GRC) to transmit standard-compliant IEEE 802.11a WiFi packets through the SDR. 
Using $\texttt{set\_iq\_balance}$ and $\texttt{set\_dc\_offset}$ functions in GRC, these two separate complex correction factors can be set to intentionally introduce required level of impairments in the radio. 


\subsection{Experimental setup for Trace Data collection}

We study the performance of the CNN using IQ samples collected from an experimental setup of USRP SDRs, as shown in Fig.~\ref{fig:dataCollect}, with a fixed USRP B210 as the receiver. All transmitters are bit-similar USRP X310 radios that emit IEEE 802.11a standards compliant frames generated via a MATLAB WLAN System toolbox. The data frames generated contain random payload but have the same address fields, and are then streamed to the selected SDR for over-the-air wireless transmission. The receiver SDR samples the incoming signals at $5\ \textrm{MS/s}$ sampling rate at center frequency of $2.45\  \textrm{GHz}$ for WiFi. The collected complex IQ samples are partitioned into subsequences. For our experimental study, we set a fixed subsequence length of $128$, i.e., the length of contiguous samples that will be used at a time for training and classification. Overall, we collect over $20$ million samples for each radio, subsequently divided into training, validation and test set.

\section{CNN architecture for static channels} \label{sec:4}
\subsection{Classifier architecture}
\begin{figure}
\centering
   \includegraphics[width=0.9\linewidth]{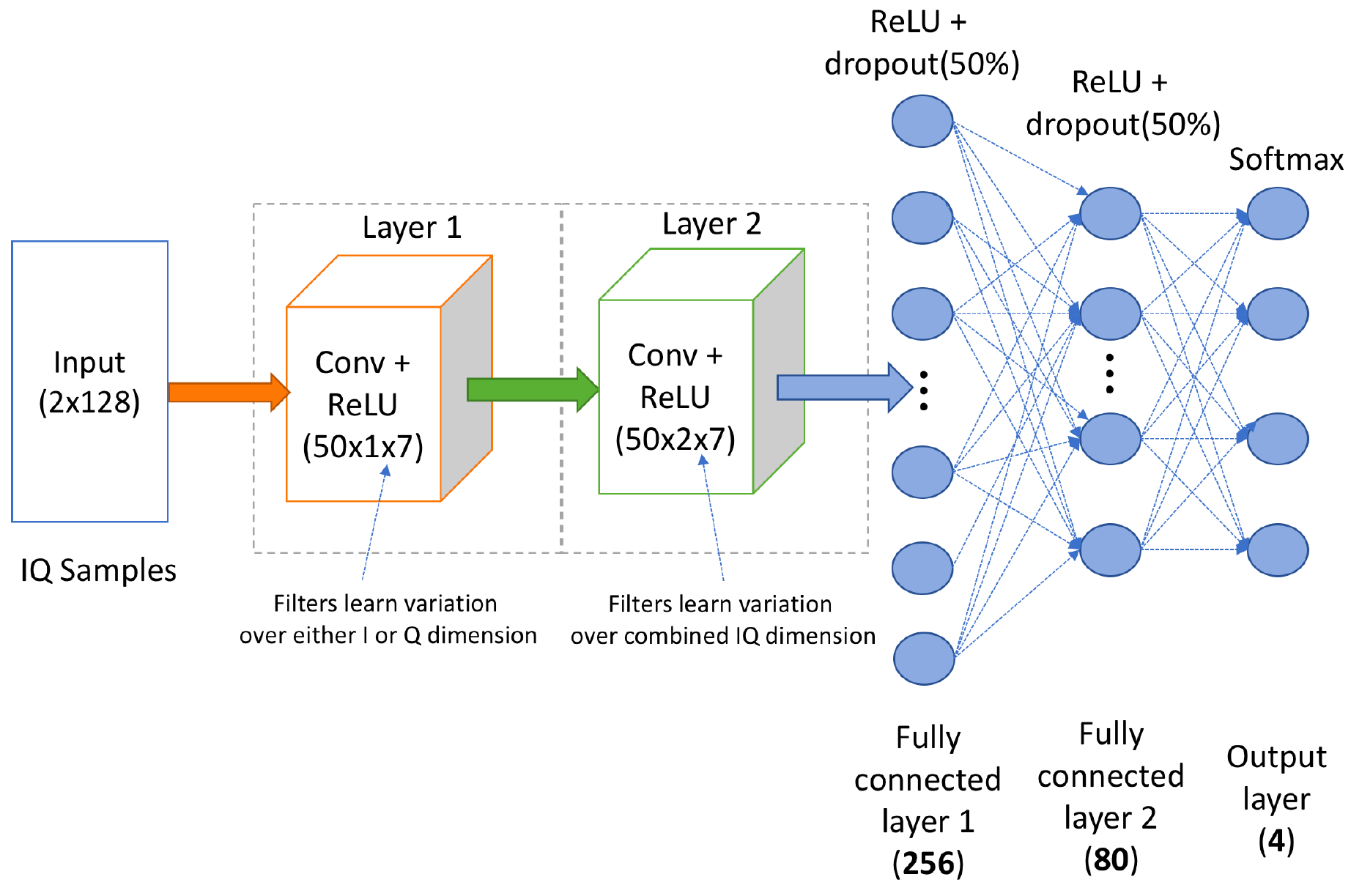}
   \caption {Our proposed CNN architecture with two convolution and two fully connected layers. }
   \label{fig:CNNarchi}\vspace{-0.5cm}
\end{figure}
For static channels, we design a CNN architecture that uses raw time-series IQ samples generated from 16-node USRP X310 SDR testbed and the external database of $140$ COTS WiFi devices. Our proposed CNN architecture, as shown in Fig.~\ref{fig:CNNarchi}, is partly inspired from AlexNet \cite{AlexNet}. It is a deep CNN architecture specifically designed to classify 1.2 million high-resolution images available in the ImageNet dataset into 1000 different classes. Unlike AlexNet, which is made up of 8 layers (5 convolution and 3 fully connected), our CNN architecture consists of four layers, with two convolution layers and two fully connected (or dense) layers. The input to our CNN is a windowed sequence of raw IQ samples with length 128. We choose a \textit{sliding window} approach to partition the training samples that enhances the shift invariance of the features learned by the CNN. Each complex value is represented as two-dimensional real values (i.e., I and Q are two real value streams), which results in the dimension of our input data growing to $2 \times 128$. This is then fed to the first convolution layer. The convolution layer consists of a set of spatial filters, also called \textit{kernels}, that perform a convolution operation over input data to extract the features. The first convolution layer consists of 
50 filters, each of size $1 \times 7$, in which each filter learns a 7-sample variation in time over the I or Q dimension separately, to generate 50 distinct feature maps over the complete input sample. Similarly, the second convolution layer has 50 filters each of size $2 \times 7$ and each filter learns variations, again of 7 activation values, over both I and Q dimensions of the 50-dimensional activation volume obtained after the first convolution layer.  Each convolution layer is followed by a Rectified Linear Unit (ReLU) activation, that performs a pre-determined non-linear transformation on each element of the convolved output.



The output of the second convolution layer is then provided as input to the first fully connected layer, which has 256 neurons. A second fully connected layer of 80 neurons is added to extract higher level non-linear combinations of the features extracted from previous layers, which are finally passed to a classifier layer.
A \textit{softmax} classifier is used in the last layer to output the probabilities of each sample being fed to the CNN.  
The choice of hyperparameters such as filter size, number of filters in the convolution layers and the depth of the CNN is of high importance to ensure that our CNN model generalizes well. These are chosen carefully through cross validation. 
In order to overcome overfitting, we set the dropout rate to 50\% at the dense layers. We also use an $\ell_2$ regularization parameter $\lambda=0.0001$. The weights of the network are trained using Adam optimizer with a learning rate of $lr = 0.0001$. We minimize the prediction error through back-propagation, using categorical cross-entropy as a loss function computed on the classifier output. We implement our CNN architecture in Keras running on top of TensorFlow on a system with 8 NVIDIA Cuda enabled Tesla K80m GPU. 

\subsection{Preliminary results}
\begin{figure}
  \centering
  \includegraphics[width=2.2in]{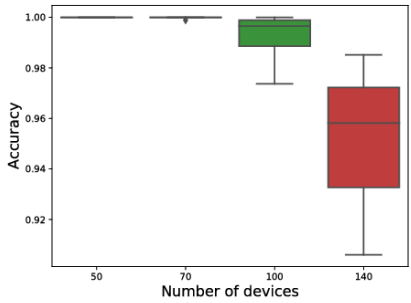}
   \caption{\small \label{fig:WiFi_ExtDatabase} Box plot for the classification of WiFi devices using CNN.}\vspace{-0.5cm}
\end{figure}
Our preliminary evaluation aims to demonstrate the accuracy of ORACLE's CNN architecture for classifying radios under static conditions, and it also motivates the need for receiver-feedback driven modifications for dynamic channels using techniques described in Sec.~\ref{subsubsec:utilities}. 

\subsubsection{Accuracy in static channel conditions}
First, we verify the performance of our proposed CNN to classify COTS WiFi devices using an external database, which contains labeled IQ samples collected from 140 devices (phones/tablets/laptops/drones) of 122 manufacturers.  
For each device, we use 4.5K windowed  examples as training set and 1K examples as test set, based on available samples in the database. A validation set of 300 examples for each device is used at each training epoch to monitor the performance on unseen data and the training process is stopped if the validation accuracy does not increase for 10 consecutive training epochs. The training time for this experiment using all 140 devices is $\approx 15min$.
ORACLE's performance is shown in Fig.~\ref{fig:WiFi_ExtDatabase} with the minimum accuracy, first quartile, median, third quartile, and maximum accuracy for each dataset. Here, the X-axis represents a number of randomly chosen devices whereas the classification accuracy is shown on the Y-axis. 
Up to 100 different devices, we obtain a median accuracy of 99\%, whereas it is 96\% for 140 devices. We note that while the number of radios is large, these devices are not bit-similar.   
Hence, we `stress-test' our classifier using collected IQ samples from 16, high-end X310 USRP SDRs that present a narrower range of impairments, with the same B210 radio as a receiver. Our training set for this experiment consists, per radio, of $200K$ windowed training examples and $10K$ examples for validation. We use another $50K$ examples for each device to test the performance of our trained model.  It takes $\approx 30min$ with our current setup to train the model for 16 radios. 
Also for this setup, we obtained 98.6\% accuracy on the test set, shown in Fig.~\ref{fig:confnorm_8_good}. 

\subsubsection{Limitations of raw IQ samples in dynamic channels}



\begin{figure}[!t]
    \centering
    \begin{subfigure}{0.25\textwidth}
    \centering
        \includegraphics[width=\linewidth]{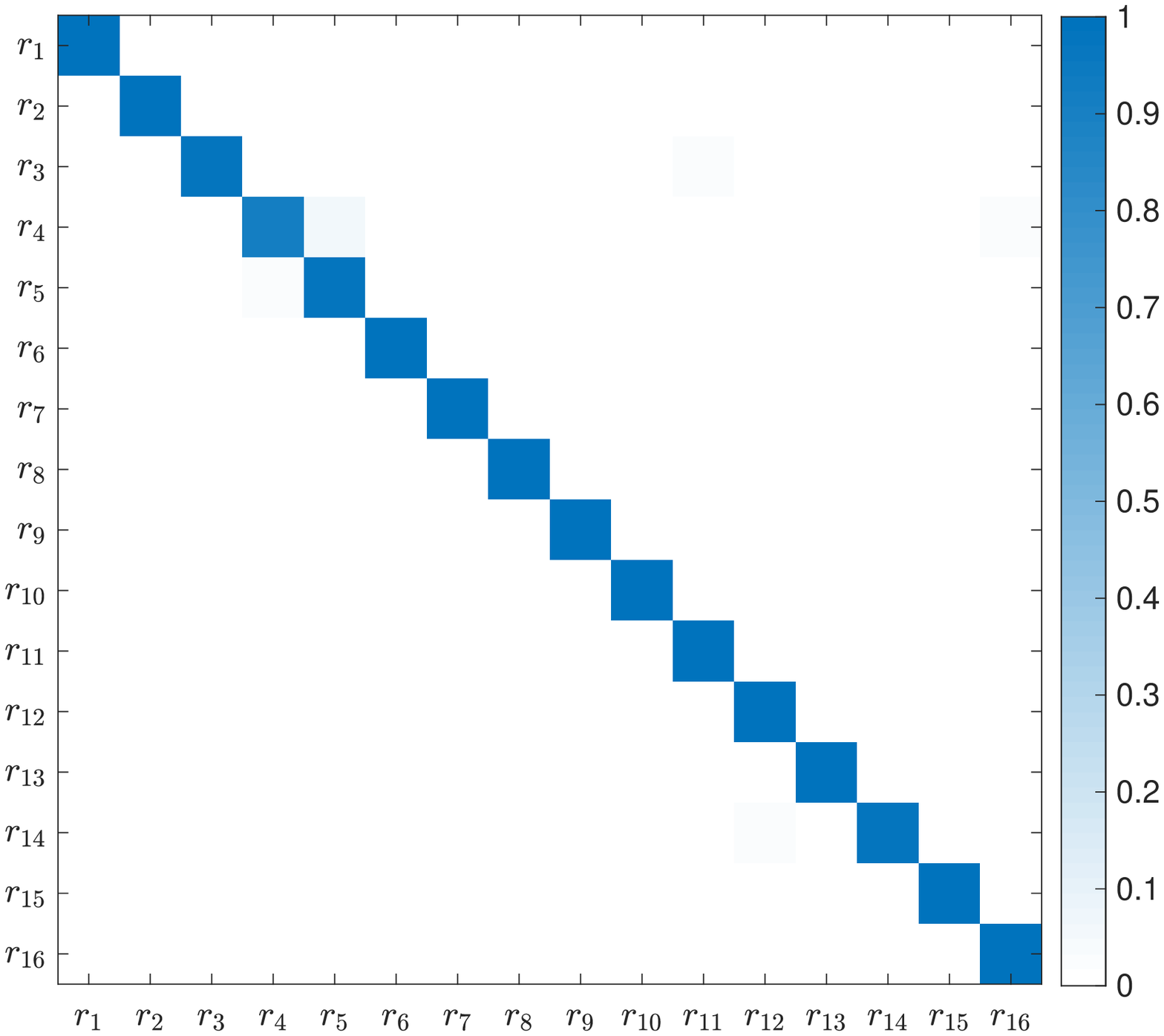}
        \caption{}
        \label{fig:confnorm_8_good}
    \end{subfigure}%
    \begin{subfigure}{0.25\textwidth}
    \centering
        \includegraphics[width=\linewidth]{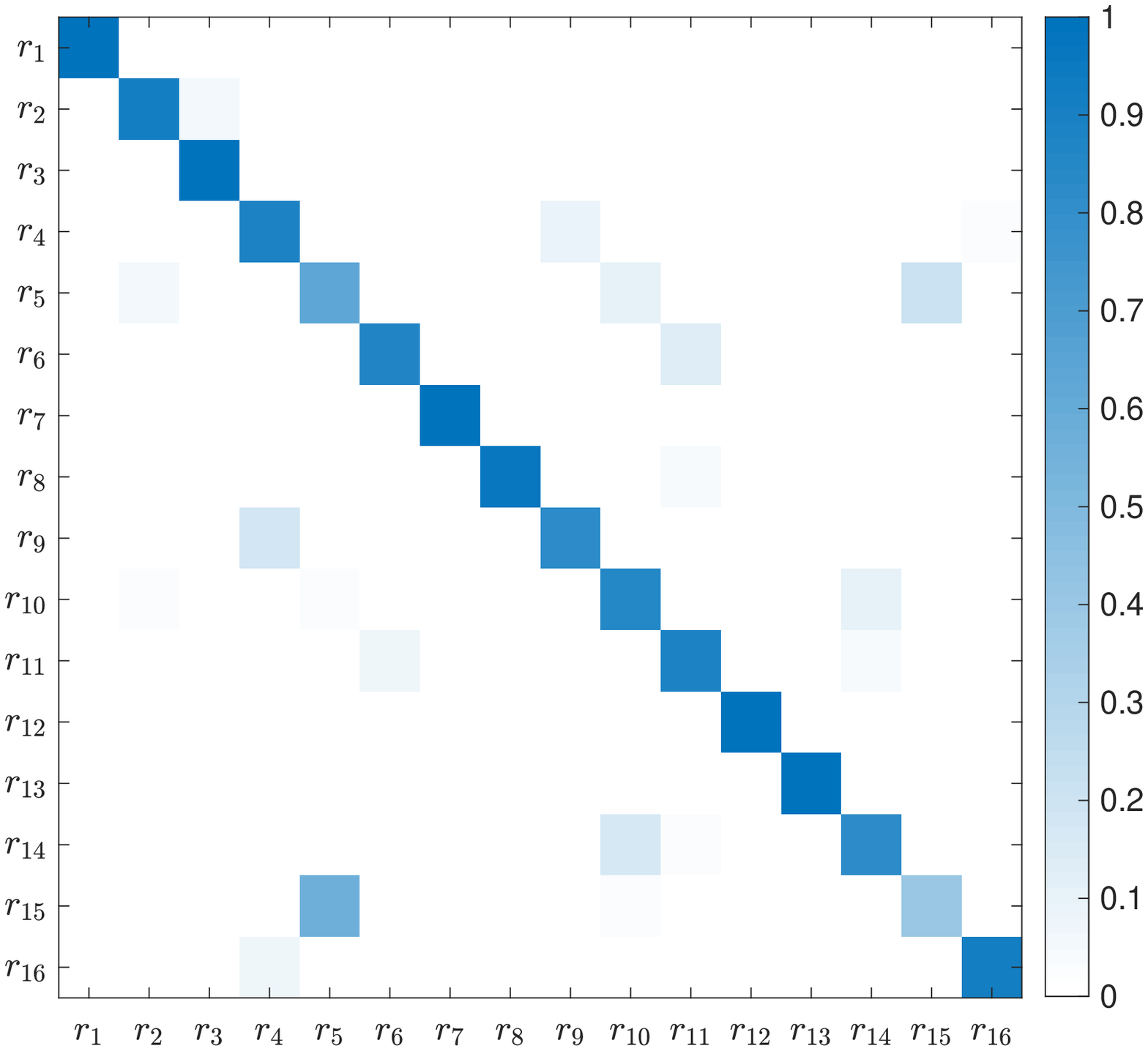}
        \caption{}
        \label{fig:confnorm_8_lessgood}
    \end{subfigure}
    \caption{ Confusion matrix relative to two experiments with same devices and different locations: (a) overall accuracy is $98.60\%$; (b) overall accuracy is $87.13\%$. }\vspace{-0.3cm}
    \label{fig:16dev_performance_prem}
    \vspace{-0.2cm}
    \end{figure}


\begin{figure}[!t]
    \centering
    \begin{subfigure}{0.25\textwidth}
    \centering
        \includegraphics[width=\linewidth]{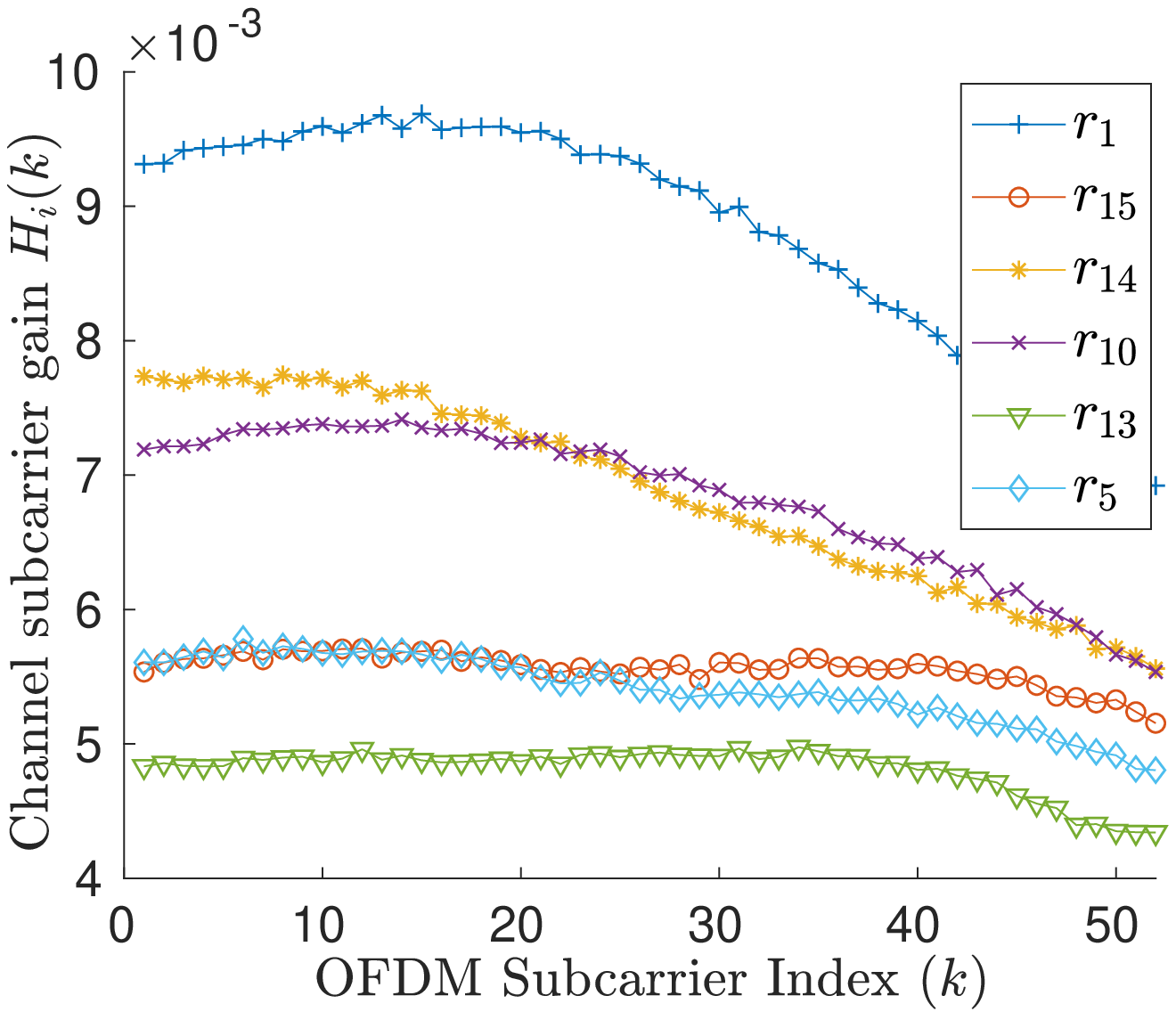}
        \caption{}
        \label{fig:chnl_est}
    \end{subfigure}%
    \begin{subfigure}{0.25\textwidth}
    \centering
        \includegraphics[width=\linewidth]{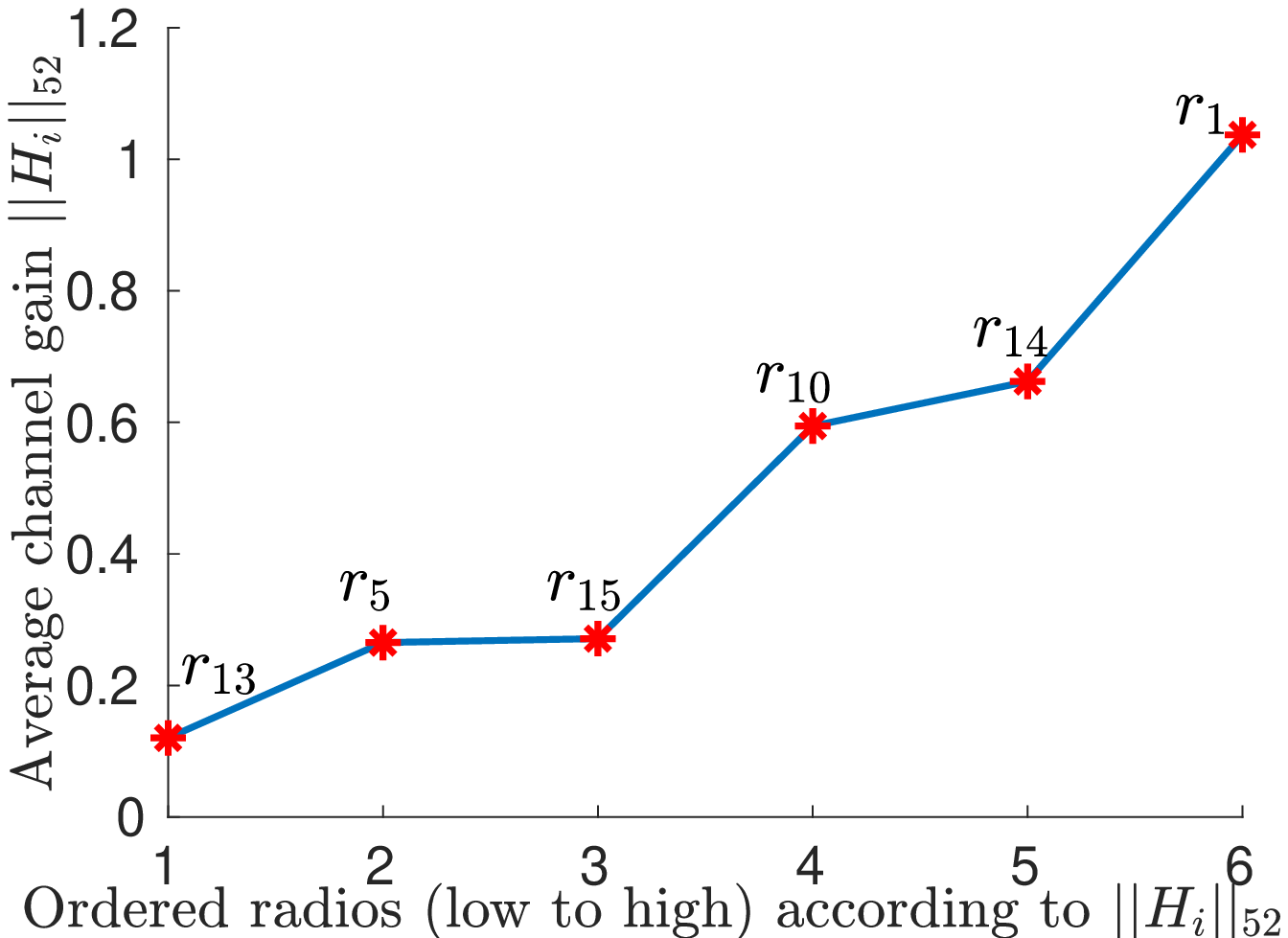}
        \caption{}
        \label{fig:chnl_magnitude}
    \end{subfigure}
    \caption{(a) Estimated channel gain $\tilde{H}_i(k)$ for $k^{th}$ subcarrier for each radio $r_i \in R$ (b) Magnitude of estimated channel $||\tilde{H}_i||_{52}$ for all radios $r_i \in R$ (ordered from lower to higher). }\vspace{-0.5cm}
    \label{fig:EMDconfusion}
    \end{figure}


Multipath reflection and fading have considerable impact on received IQ samples, at times distorting the samples wherein the classifier no longer correctly identifies the radios. Typically, the effect of the channel is compensated by channel estimation and equalization techniques to correctly retrieve over-the-air transmitted data. 
Thus, as we show next, classification performance degrades severely when either (i) classifiers are trained on raw IQ samples under a given channel and then tested on IQ samples obtained under different channels, or (ii) transmitters experience very similar channel conditions.



Fig.~\ref{fig:confnorm_8_good} shows the classification accuracy of 16 X310 radios, with near-perfect results for all the devices. 
However, Fig.~\ref{fig:confnorm_8_lessgood} shows the same setup in a different location where several outliers exist, as the confusion matrix shows, e.g., see radio pairs (5,15), (10, 14). The reason is that \emph{the similarity in the  wireless channel experienced by certain transmitter pairs dominates subtle hardware variations}. 
Given a set of $R$ radios, $\tilde{H}_i(k)$ represents the average channel gain in $k^{th}$ subcarrier of each radio $r_i \in R$, estimated over WiFi packets belonging to the training dataset.

\begin{figure}[!t]
    \centering
    \begin{subfigure}{0.16\textwidth}
    \centering
        \includegraphics[width=\linewidth]{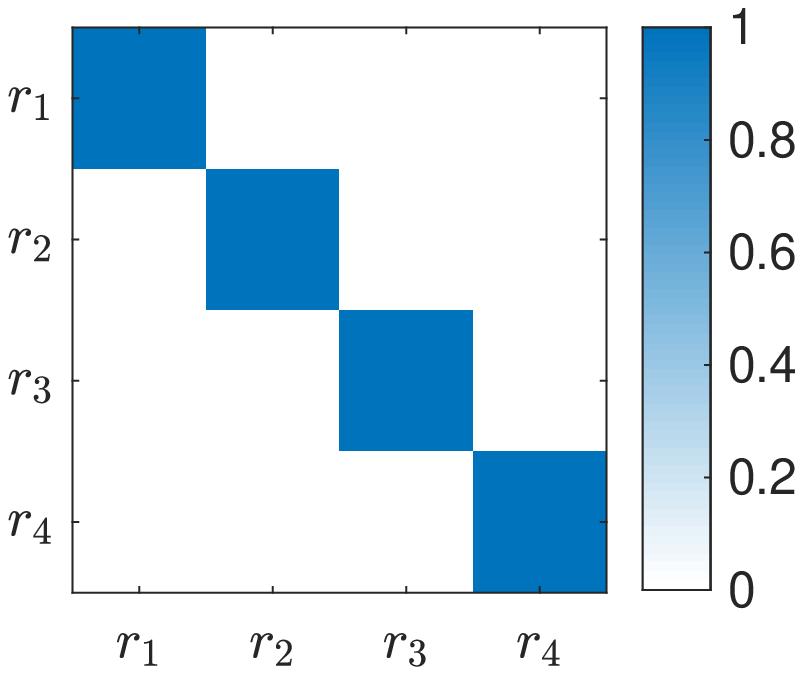}
        \caption{}
        \label{fig:orig_4dev}
    \end{subfigure}%
    \begin{subfigure}{0.16\textwidth}
    \centering
        \includegraphics[width=\linewidth]{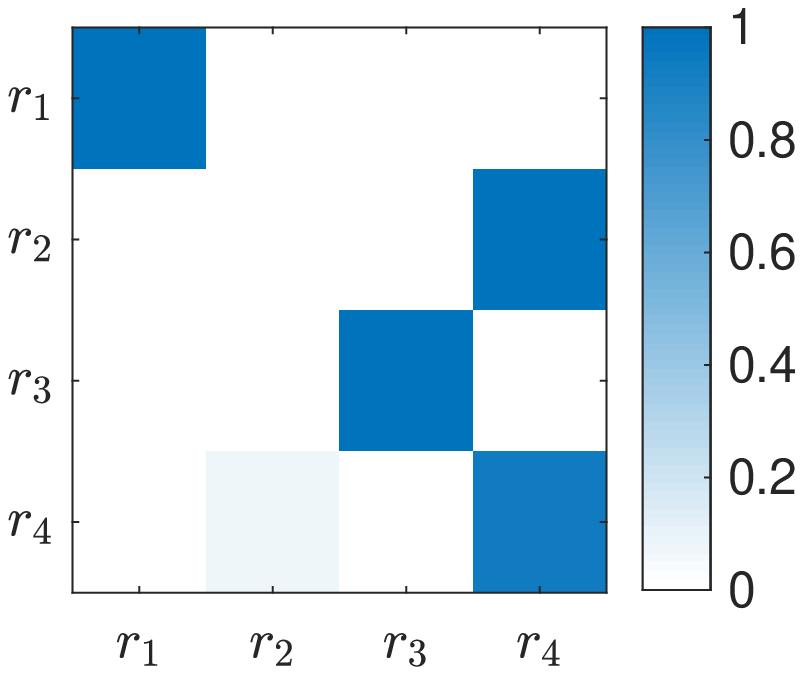}
        \caption{}
        \label{fig:time2_4dev}
    \end{subfigure}
     \begin{subfigure}{0.16\textwidth}
    \centering
        \includegraphics[width=\linewidth]{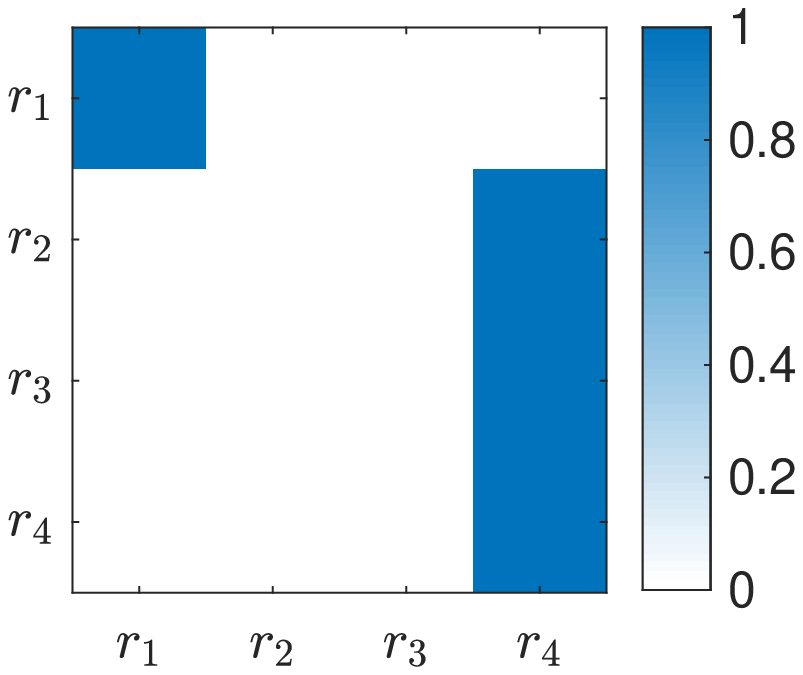}
        \caption{}\label{fig:location2_4dev}
    \end{subfigure}
    \caption{(a) Classification accuracy for 4 devices tested at time $t_1$ and location $l_1$; (b) time $t_2$ and same location $l_1$; (c) time $t_3$ and different location $l_3$. }\vspace{-0.5cm}
    \label{fig:EMDconfusion}
    \end{figure}

Fig.~\ref{fig:chnl_est} and \ref{fig:chnl_magnitude} reveal how received samples from transmitters with  smaller differences in channel estimation are more likely to be misclassified by ORACLE during testing. 
This shows that wireless channel state affects the distribution of complex symbols captured by the receiver in a non-negligible manner, and therefore \emph{becomes a discriminating factor when the classifier is trained with raw IQ samples}.
If we try to use a pre-trained model and use it to classify samples collected from same devices but at different times or locations, the classification result is unpredictable. See Fig.~\ref{fig:orig_4dev}, \ref{fig:time2_4dev} and \ref{fig:location2_4dev} for the classification results showing the time and location dependence of the trained classifier. 

\section{ORACLE with Feedback for Dynamic Channels}
This section describes the enhancements in ORACLE that allow it to robustly classify transmitters in unseen environments. The two main assumptions here are: (i) instead of raw IQ samples, ORACLE works with demodulated symbols, and (ii) in a pre-deployment phase, the receiver provides feedback to the transmitter to incorporate controlled impairments.

\subsection{Impact of impairments on demodulated symbols} \label{sec:5a}
\begin{figure}{}
  \centering
  \includegraphics[width=\linewidth]{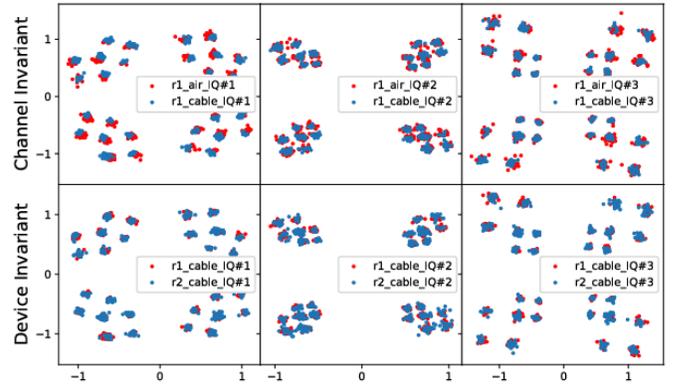}
   \caption{Patterns generated by 3 impairments on 2 devices under 2 channel conditions. First and second row show the channel- and device- invariance of the patterns respectively.} \label{fig:pattern}
   \vspace{-0.5cm}
\end{figure}

ORACLE modifies the transmitter chain of the SDRs such that their respective demodulated symbols acquire unique characteristics that make the CNN robust to channel changes, i.e., it makes the transmitter hardware \textit{dominate} channel induced variations. We first validate the hypothesis that a given combination of impairments results in repeatability in the outcome of the classification. 
To demonstrate this, consider demodulated symbols received from two X310 radios, over cable and air channels, as shown in Fig.~\ref{fig:pattern}, for three different levels of IQ imbalance. The first row shows slight differences in the demodulated samples when the channel is completely changed (i.e., air to cable) for the same transmitter. In the second row, when the same channel is maintained, but the transmitters themselves are different, adding the same level of IQ imbalance results in virtually the same pattern in each case, ensuring repeatability and robustness.


We also quantitatively analyze the property of the channel- and device- invariance of the patterns with Earth Mover's Distance (EMD), a widely used metric to measure similarities between two multi-dimensional distributions. 
More precisely, suppose we have two sets of points in $\mathbb{R}^2$. Let $A \subset \mathbb{R}^2$ and $B \subset \mathbb{R}^2$ be two subsets of equal size, i.e., $|A| = |B|$. Let $F$ be the set of all possible bijections ($1-1$ and onto mappings) from $A$ to $B$. The EMD between $A$ and $B$ is given by: 

\begin{equation} \label{eq:emd}
EMD(A, B) = \min_{f \in F} \sum_{x \in A} \lVert x - f(x) \rVert.
\vspace{-1mm}
\end{equation}

In other words, EMD is given by the smallest possible sum of Euclidean distances between points in $A$ and $B$, over all possible valid bijections $f:A \rightarrow B$. Smaller EMD indicates more similarities between two patterns and vice versa. Fig.~\ref{fig:EMDconfusion} (a) and (b) show the EMD matrix of patterns generated on different channel conditions and devices respectively with the same set of impairments in Fig.~\ref{fig:pattern}. We see that computed EMD on the matrix diagonal, which represents the patterns generated by the same impairments, are much lower than the EMD of patterns generated by different impairments. We further evaluate the EMD for the demodulated signal collected under 3 different channel conditions, 4 devices across 32 different levels of impairments. We see that the average EMD remains around 0.1 and 0.2 for patterns generated by the same and different level of impairments, respectively, despite of the variations caused by channel conditions. This result matches closely with Fig.~\ref{fig:EMDconfusion} and verifies our intuition.

\begin{figure}[!t]
    \centering
    \begin{subfigure}{0.24\textwidth}
    \centering
        \includegraphics[width=\linewidth]{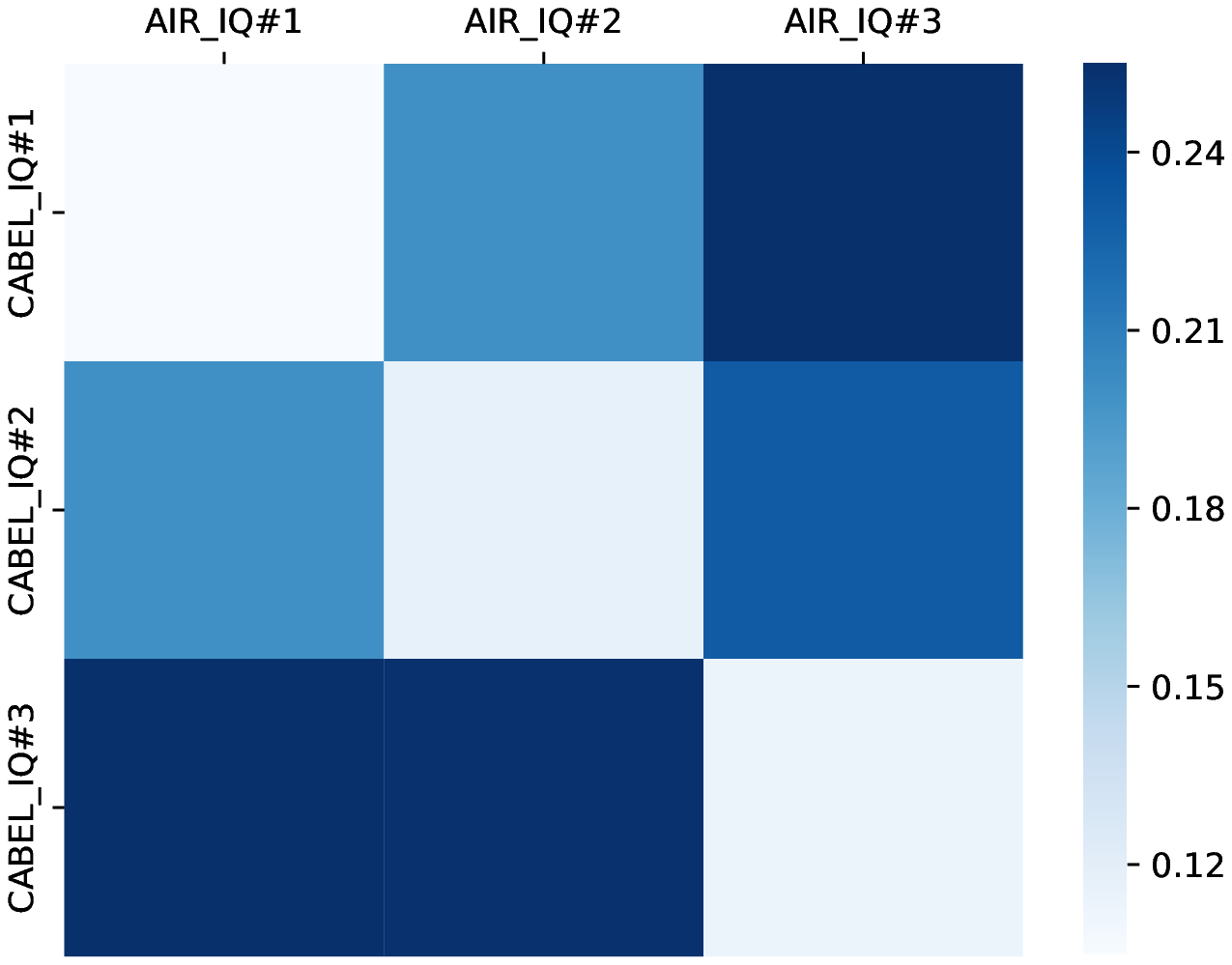}
        \caption{}
        \label{fig:EMDconfusion1}
    \end{subfigure}%
    \begin{subfigure}{0.24\textwidth}
    \centering
        \includegraphics[width=\linewidth]{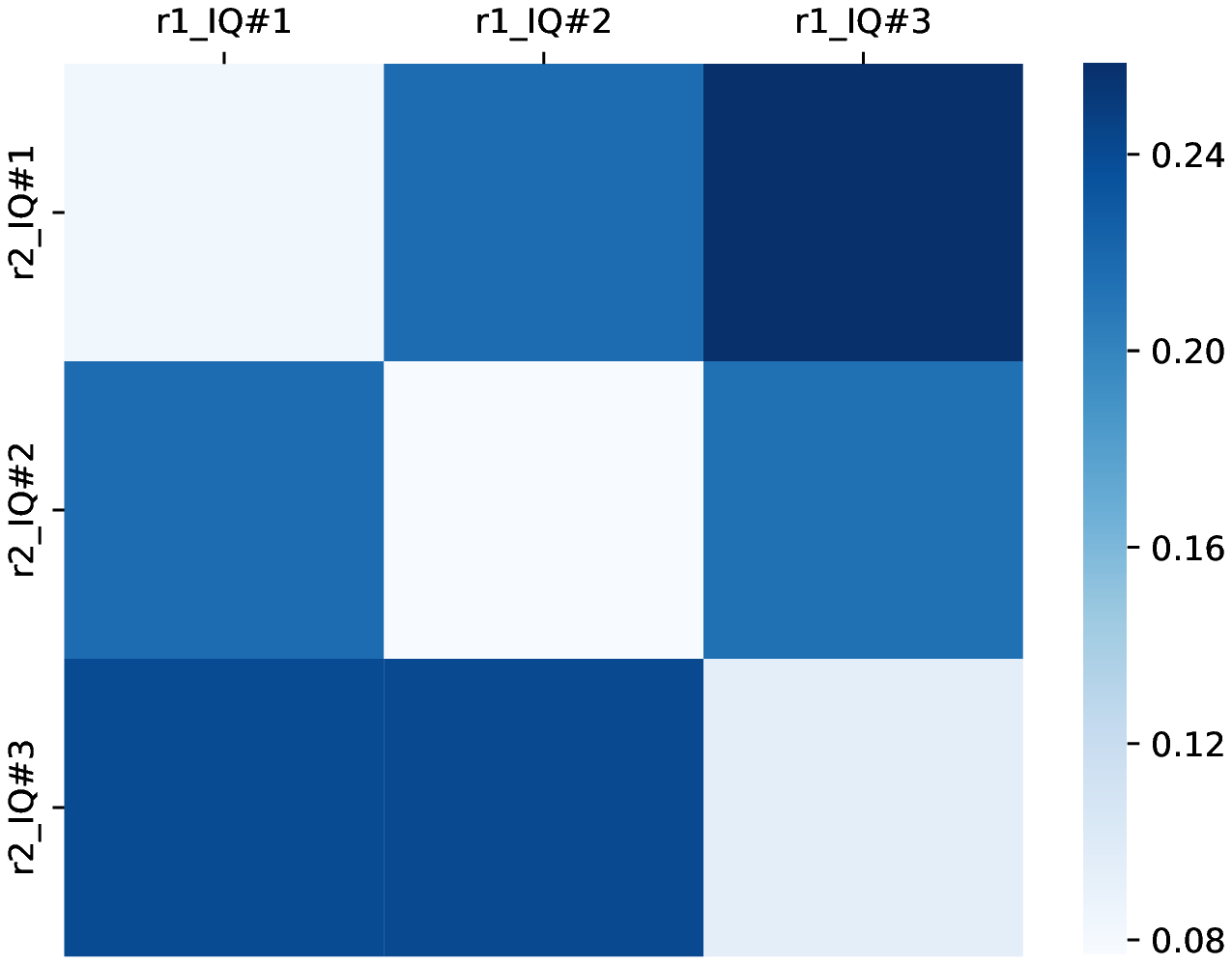}
        \caption{}
        \label{fig:EMDconfusion1}
    \end{subfigure}
    \caption{The EMD matrix of patterns generated (a) under different channel conditions; (b) on different devices. \vspace{-0.5cm}}
    \label{fig:EMDconfusion}
    \end{figure}

\subsection{Identifying feasible impairments} \label{sec:5b}

\label{feasible_imp}

The naive approach of introducing random combinations of impairments before training the CNN has three problems: 
\begin{enumerate}
\item \emph{Scalability:} If a new transmitter is introduced in the network, then we have to re-train the entire CNN, which is a time- and computation-heavy process. 
\item \emph{Accuracy:} It is possible that demodulated samples originating from two different transmitters (previously, easily differentiable) now appear clustered together owing to the modification in their placement on the IQ plane. This may reduce the performance of the classifier. 
\item \emph{Communication impact:} Adding impairments naturally increases the BER. Hence judicious and controlled addition is needed to limit any adverse impact on BER.
\end{enumerate}

To solve these issues, ORACLE automatically selects feasible impairments that produce IQ sample constellation points that are significantly different from each other, while minimizing the influence on the BER for the transmitter. This step allows ORACLE to pre-train on virtual radios transmitter chains (constructed in GNU Radio) as the impairments dominate other variations introduced by its own hardware and the wireless channel. Thus, ORACLE learns the impairment patterns, which we have shown in Fig.~\ref{fig:pattern} to be  both device and channel agnostic, i.e., two different radios will result in a similar demodulated IQ pattern at the receiver under the same impairment. 
This approach greatly increases the flexibility of ORACLE: if a new transmitter is added, we simply assign it one of the feasible and uncommitted impairments, without any need to re-train the CNN. 

\begin{figure}[!t]
    \centering
    \begin{subfigure}{0.25\textwidth}
    \centering
        \includegraphics[width=\linewidth]{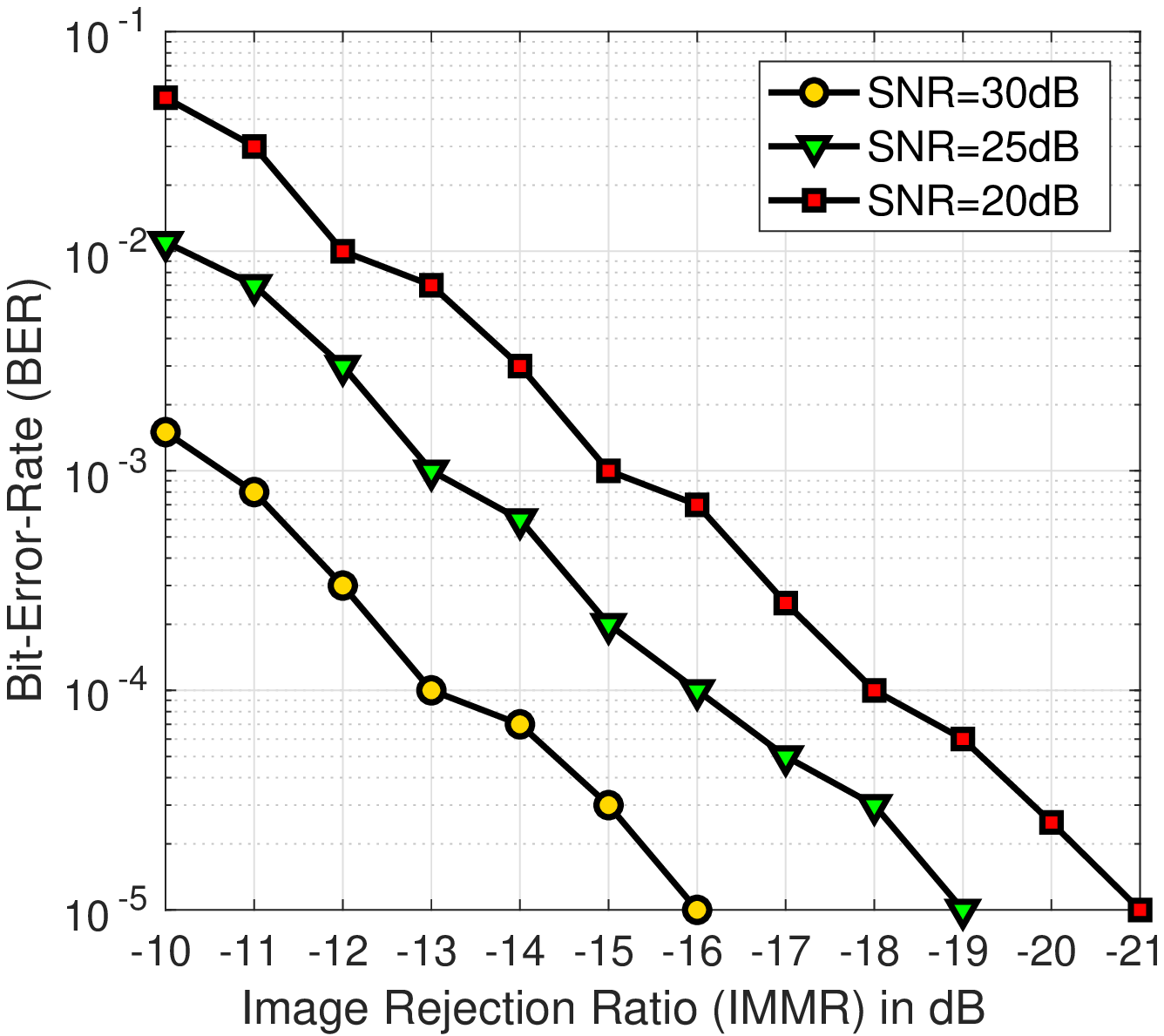}
        \caption{}
        \label{fig:BERvsIQ}
    \end{subfigure}%
    \begin{subfigure}{0.25\textwidth}
    \centering
        \includegraphics[width=\linewidth]{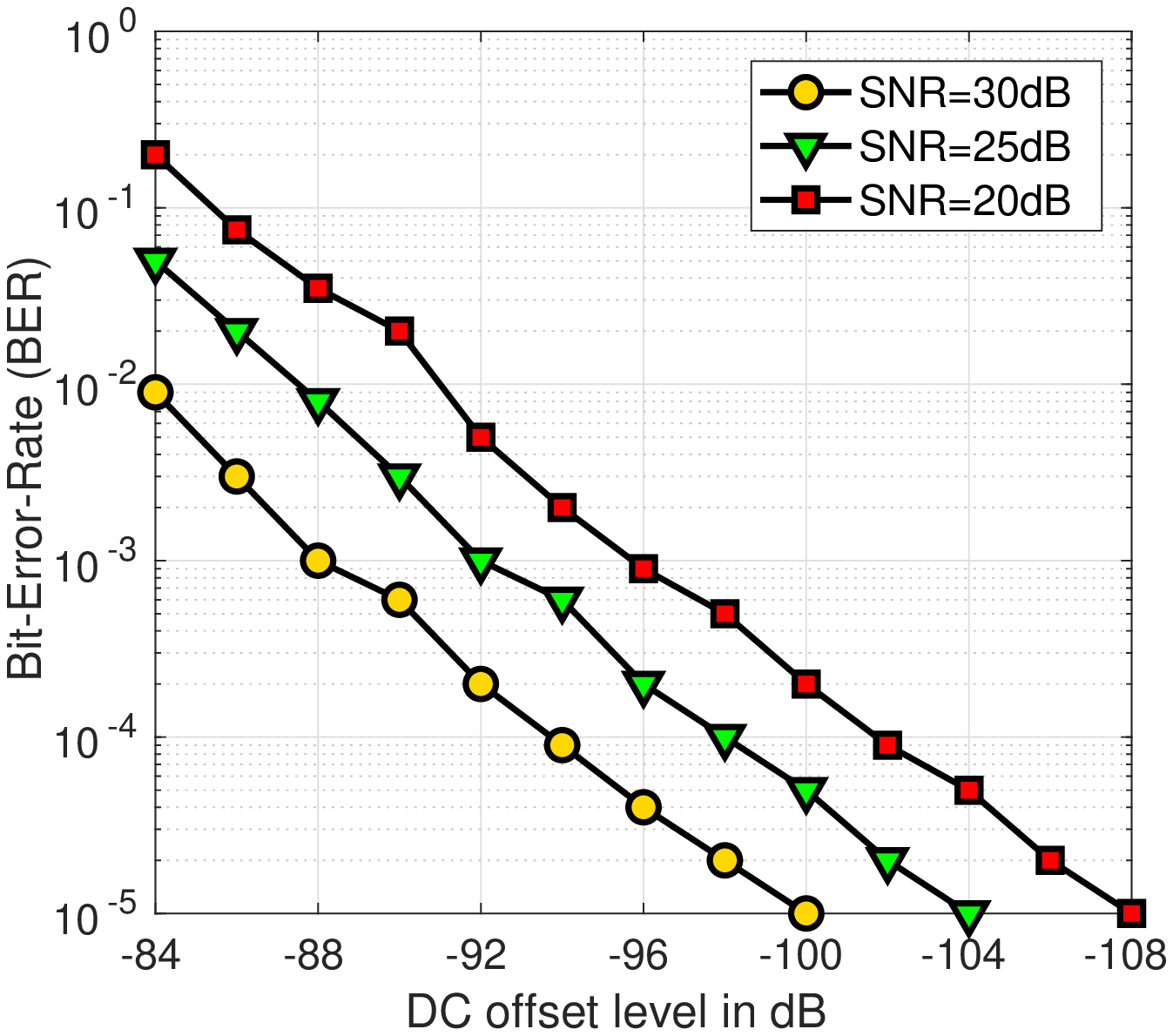}
        \caption{}
        \label{fig:BERvsDC}
    \end{subfigure}
    \caption{(a) BER vs. IMMR value of IQ imbalance; (b) BER vs. DC offset level for different SNRs.}\vspace{-0.5cm}
    \label{fig:BER}
    \end{figure}

We use a generic X310 USRP radio that operates in a loop while automatically adding impairments to its hardware through the utilities $\texttt{uhd\_cal\_tx\_iq\_balance}$ and $\texttt{uhd\_cal\_tx\_dc\_offset}$ for IQ imbalance and DC offset, respectively. 
Then the transmitter sends a stream of known data over cable to the B210 USRP receiver that checks the BER. 
For our experiment, we consider $80$ different levels of IQ imbalance with IMMR value ranging from $-9$~dB to $-44$~dB and $120$ levels of DC offset ranging from $-82$~dB to $-140$~dB. The BER plots are shown in Fig.~\ref{fig:BERvsIQ} and Fig.~\ref{fig:BERvsDC} for different SNR levels, which we concisely refer to as an impairment map $M$, and use it  later in Sec.~\ref{sec:5d}. The bounds on the impairments depend on the SNR that the radios operate in. For e.g., our lab has a noise floor of $-70$~dBm, for which we assume an average $30$~dB SNR level with the constraint on BER of $10^{-4}$. Accordingly, we choose upper bound $-13$~dB on IMMR for IQ imbalance and $-94$~dB for DC offset level. 



We next explain how to identify the feasible set $S$ out of all impairment combinations that satisfy the BER constraint. Specifically, let $[c_1, c_2, \cdots,c_{max}]$ be the vector of different levels of IQ imbalance resulting in an ordered set of corresponding BER, i.e., $BER[c_i] < BER[c_{i+1}]$. Therefore, $c_{max}$ is the maximum IQ imbalance we can add without exceeding the BER constraint. Note that the BER constraint of $10^{-4}$ is evaluated under \emph{ideal} SNR level ($40$~dB). We start from $c_1$, since it has the smallest impact on the communication, increasingly adding $c_2$ to $c_{max}$ to the set $S$. However, any new $c_i$ is eligible to be added only if the difference in EMD between the pattern generated by $c_i$ and that of any existing $c_k$ in $S$ is larger than a threshold $T$. As we have seen in Sec.~\ref{sec:5a}, 
$T= 0.15$ allows for an acceptable buffer in evaluating how close a given IQ pattern is to another. After we have reached $c_{max}$, we configure the radio with a different type of impairment until $\|S\| > N$, where $N$ is the number of bit-similar radios. 

\subsection{CNN classifier using transmitter-side impairments}
\label{sec:5c}
In this section, we discuss to train the classifier for the patterns (see Sec.~\ref{sec:5b}). 
We reuse the same CNN architecture and the input data format as described in Sec.~\ref{sec:4}. Note all IQ samples for training are collected over the cable, i.e, we remove the influence of wireless channel so that CNN can capture the pattern generated solely by hardware impairments. 

\begin{figure}[!t]
    \centering
    \begin{subfigure}{0.16\textwidth}
    \centering
        \includegraphics[width=\linewidth]{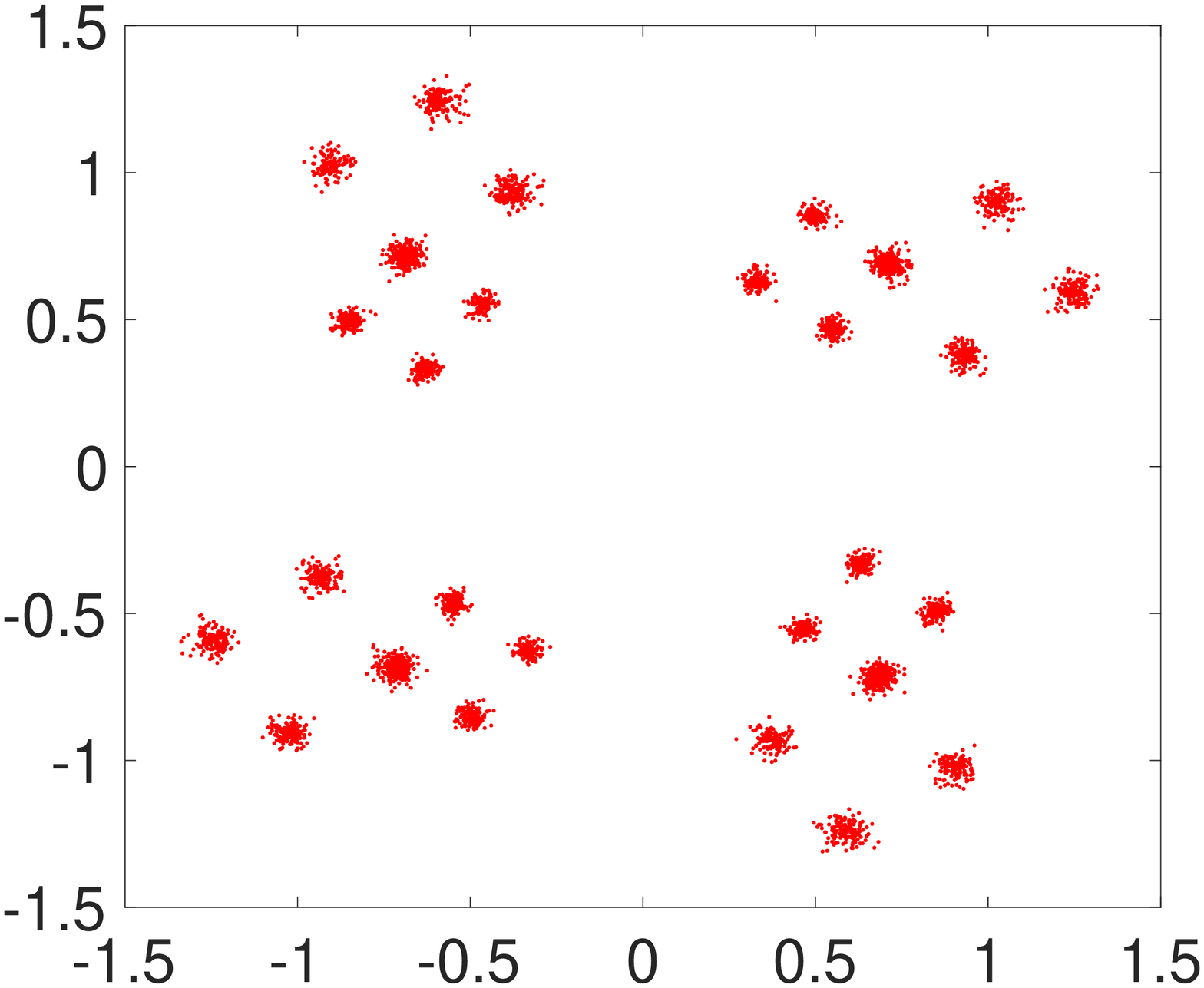}
        \caption{}
        \label{fig:no_noise}
    \end{subfigure}%
    \begin{subfigure}{0.16\textwidth}
    \centering
        \includegraphics[width=\linewidth]{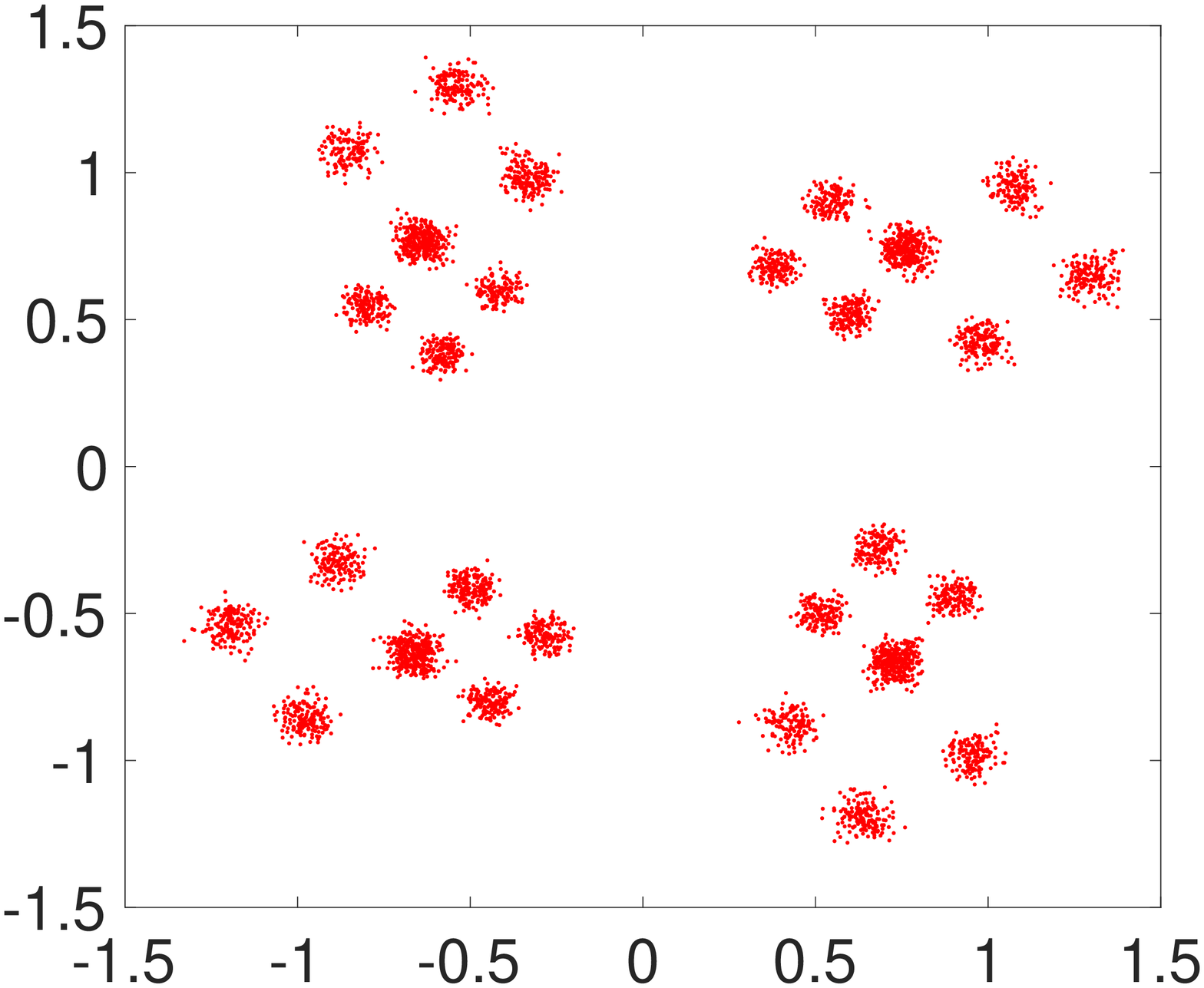}
        \caption{}
        \label{fig:lo_noise}
    \end{subfigure}
    \begin{subfigure}{0.16\textwidth}
    \centering
        \includegraphics[width=\linewidth]{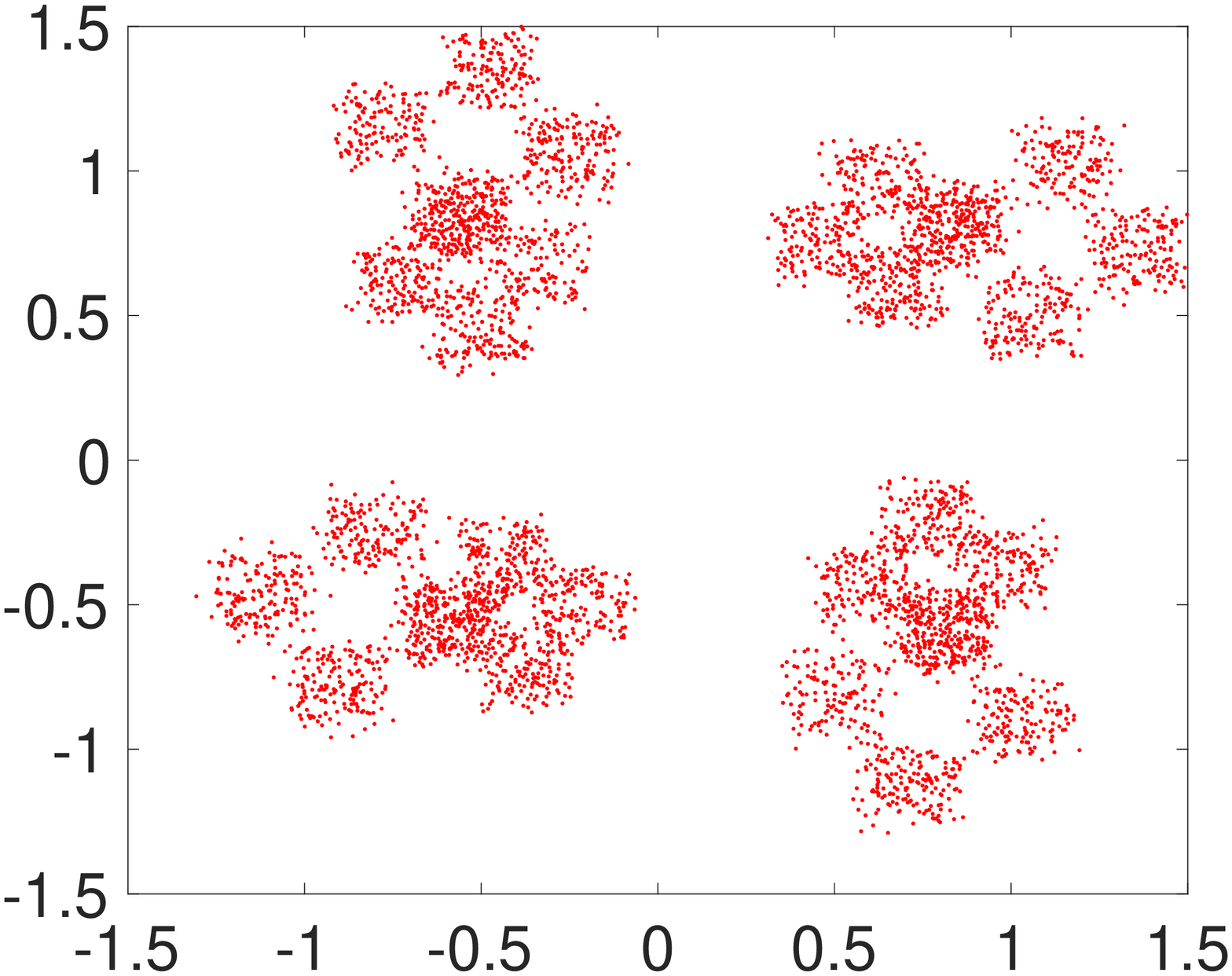}
        \caption{}
        \label{fig:hi_noise}
    \end{subfigure}
    \caption{Pattern generated with (a) original (demodulated) data; (b)  data after adding -17 dB noise, EMD with (a): 0.07; (c) data after adding -9 dB noise , EMD with (a): 0.18.}
    \label{fig:noise}
    
\vspace{-5mm}
    \end{figure}

ORACLE deliberately introduces random noise by modifying the original data to increase the number and variability of the initial dataset before input to the classifier, a technique commonly used in deep learning.  
Since low SNR of the received samples results in  scattering around the ideal constellation point location within the IQ plane, the noise is modeled as a Gaussian variable. 
We note that noise may result in an altered demodulated IQ sample pattern that is different from the original one, as shown in Fig.~\ref{fig:noise}. To finely control the possible variations, we maintain the EMD under 0.1 \textit{after} adding noise, since two sample patterns up to this level are still similar to each other (see Sec.~\ref{sec:5a}). Thus, adding noise power less than $\sigma^2_n=-13\ \textrm{dB}$ ensures that the EMD between original and altered patterns is below this threshold.

\subsection{Allocation of specific transmitters to impairments} \label{sec:5d}
\vspace{-0.1cm}
The main challenge in adding impairments is that it increases the BER and degrades the quality of service. In addition,  the degradation of impairments are different for radios under various SNR levels (as shown in Fig.~\ref{fig:BER}). Lower the SNR, the less impairments we may add to radios to ensure the required BER. We discuss how to solve this problem in this section, assuming the SNR measurements at the receiver side are quasi-static for duration $T$, allowing an average of SNR levels within each such time slot.


\noindent \textbf{Problem formulation}: Given $K$ radios $[r_1, r_2,\cdots ,r_K]$, the average SNR levels for these radios are $[snr_1, snr_2, \cdots, snr_k]$. We need to select $K$ impairments that minimize the BER of each transmitter, also depending on the average SNR level at the receiver.

We solve this problem using a greedy heuristic similar to the one we used in Sec.~\ref{sec:5a} to generate unique patterns. Without loss of generality, consider IQ imbalance with $[c_1,c_2,\cdots,c_n]$ as the set of selected IMMR levels and $M$ giving the the mapping of different SNR levels to the max IQ imbalance to maintain the BER (see Sec.~\ref{feasible_imp}). Then, for each radio $r_i$ we select $c^i_{max}$, where $c^i_{max} = M[max(Q)]$, $Q$ is the set of SNR in $M$ and $q < snr_i$,  $\forall q \in Q$. 

Following this step, we sort the radios $[r_1, r_2,\cdots,r_i]$ by their $c^i_{max}$, such that $c^i_{max} < c^{i+1}_{max}$, i.e, we sort radios according to the max IQ imbalance that can be added. Then we create two empty sets $R_1$ and $R_2$, which denotes classifiable and unclassifiable radios, respectively. We then start to allocate $[c_1, c_2,\cdots, c_n]$ iteratively to the radios from $r_1$ to $r_k$ as long as $c_i \leq c^i_{max}$ and place a given radio in the classifiable set $R_1$. Otherwise if $c_i > c^i_{max}$, it means no feasible IQ imbalance can be added to radio $i$ without exceeding the BER limit. Therefore, we put the radio in the unclassifiable set $R_2$. After we have explored all radios and if the $R_2$ is not empty, we repeat the above process with a second type of impairment (e.g., DC offset) until all radios have been put in the classifiable set.

In summary, allocating the impairment from low to high makes sure that we are minimizing the degradation in the BER. 

\section{Performance evaluation}

\begin{figure}[!t]
    \centering
    \begin{subfigure}{0.2\textwidth}
    \centering
        \includegraphics[width=\linewidth]{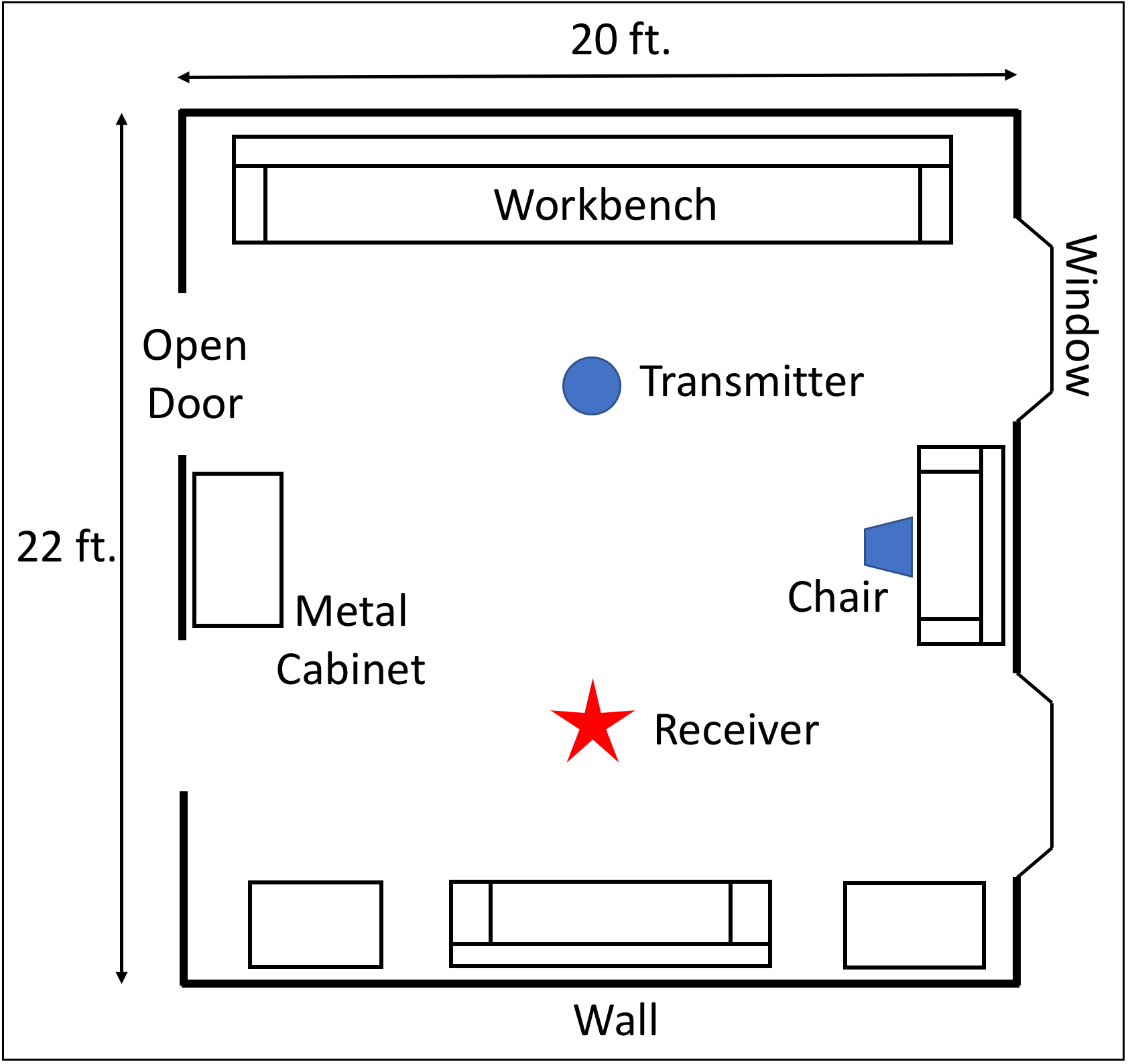}
        \caption{}
        \label{fig:location1}
    \end{subfigure}\vspace{0.01cm} %
    \begin{subfigure}{0.2\textwidth}
    \centering
        \includegraphics[width=\linewidth]{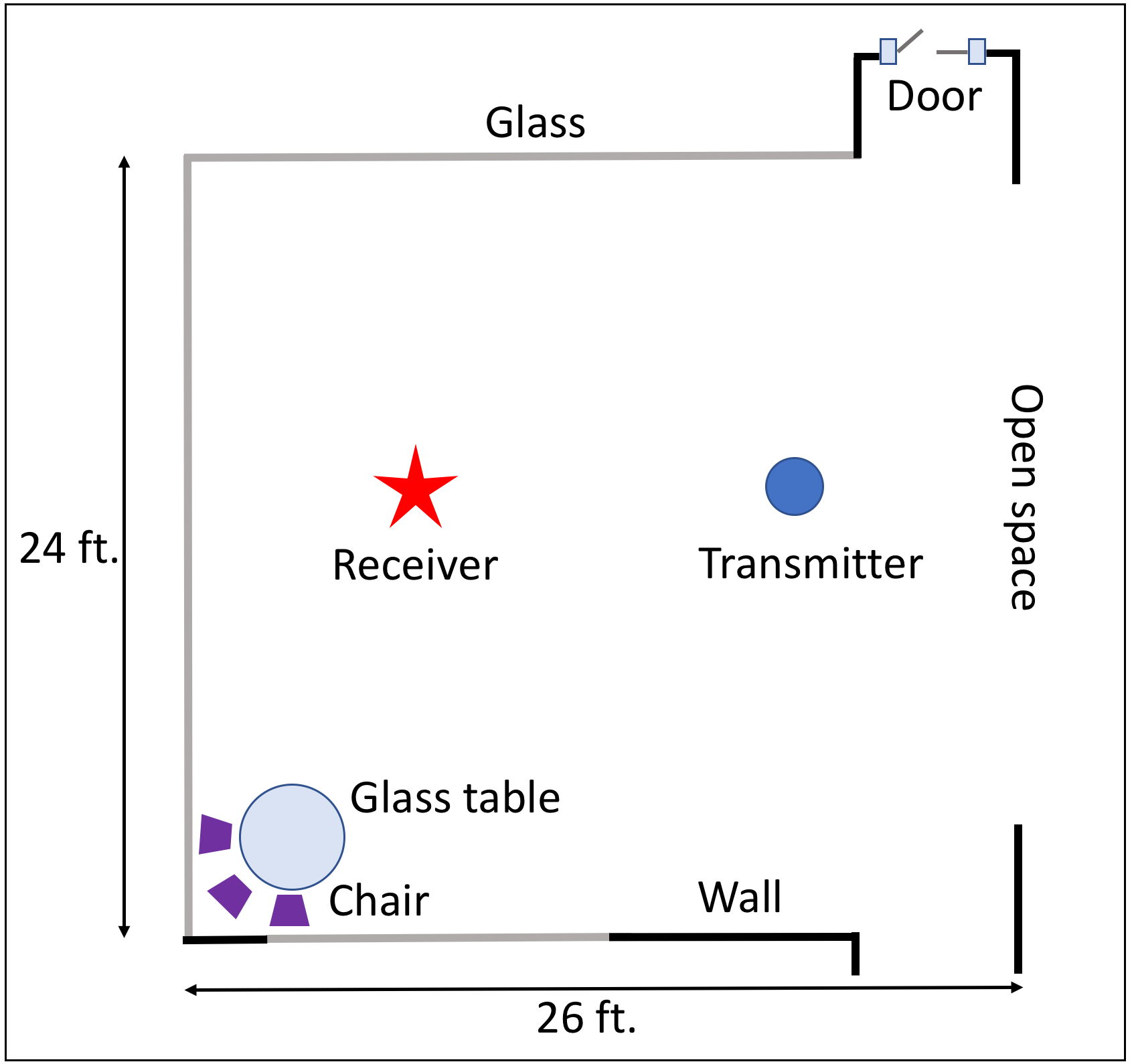}
        \caption{}
        \label{fig:location2}
    \end{subfigure}
    \caption{Two different experimental environments: (a) closed lab area (location 1); (b) open recreation area with much less reflections (location 2).}\vspace{-0.5cm}
    \end{figure}




 
 

In this section, we present the performance of ORACLE showing: (1) it increases the classification accuracy for bit-similar radios, and that accuracy is not influenced by variation in wireless channel conditions (Sec.~\ref{sec:6a}); (2) it minimizes the BER changes due to the hardware impairments without sacrificing classification performance (Sec.~\ref{sec:6b}).

\noindent \textbf{Experiment setup:} We first identify a set $S$ of 32 impairments which generates unique patterns as discussed in Sec.~\ref{feasible_imp}. Next, we collect demodulated data from WiFi packets that are transmitted over a cable from a single radio, after introducing these impairments through GNU Radio API. We replicate and augment demodulated data by adding a random Gaussian noise. We limit the power of noise to be under -13 dB to ensure that EMD lies below the threshold of $0.1$ between patterns generated from original and altered data.  Finally, we train the classifier with the augmented dataset using the same CNN architecture as described in Sec. \ref{sec:4}. 

\subsection{Classification accuracy with different channel conditions} \label{sec:6a}

\begin{figure}[!t]
    
    \begin{subfigure}{0.25\textwidth}
        \includegraphics[width=\linewidth]{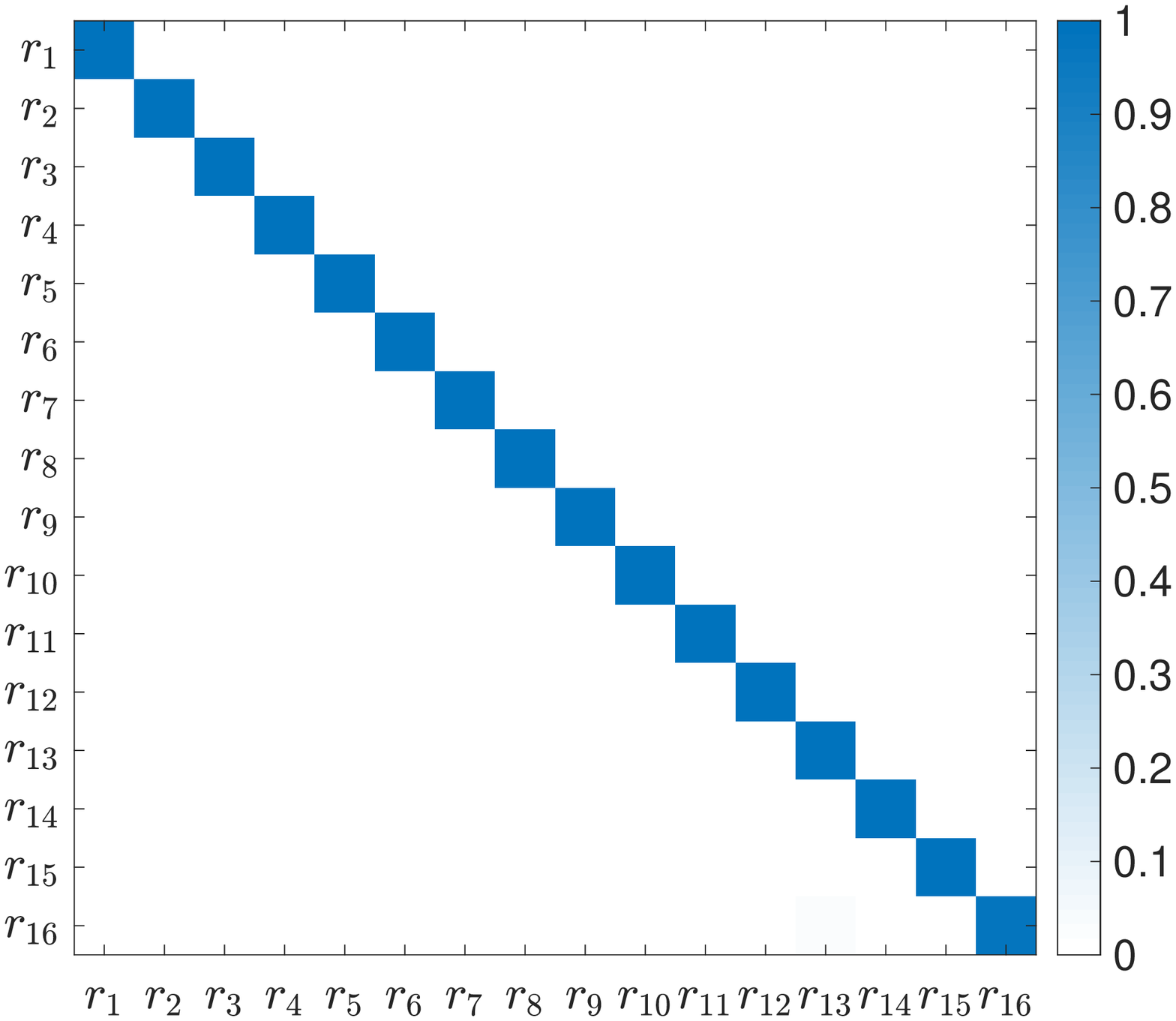}
        \caption{}
        \label{fig:4devices_Impairment}
    \end{subfigure}%
     \begin{subfigure}{0.25\textwidth}
    \centering
        \includegraphics[width=\linewidth]{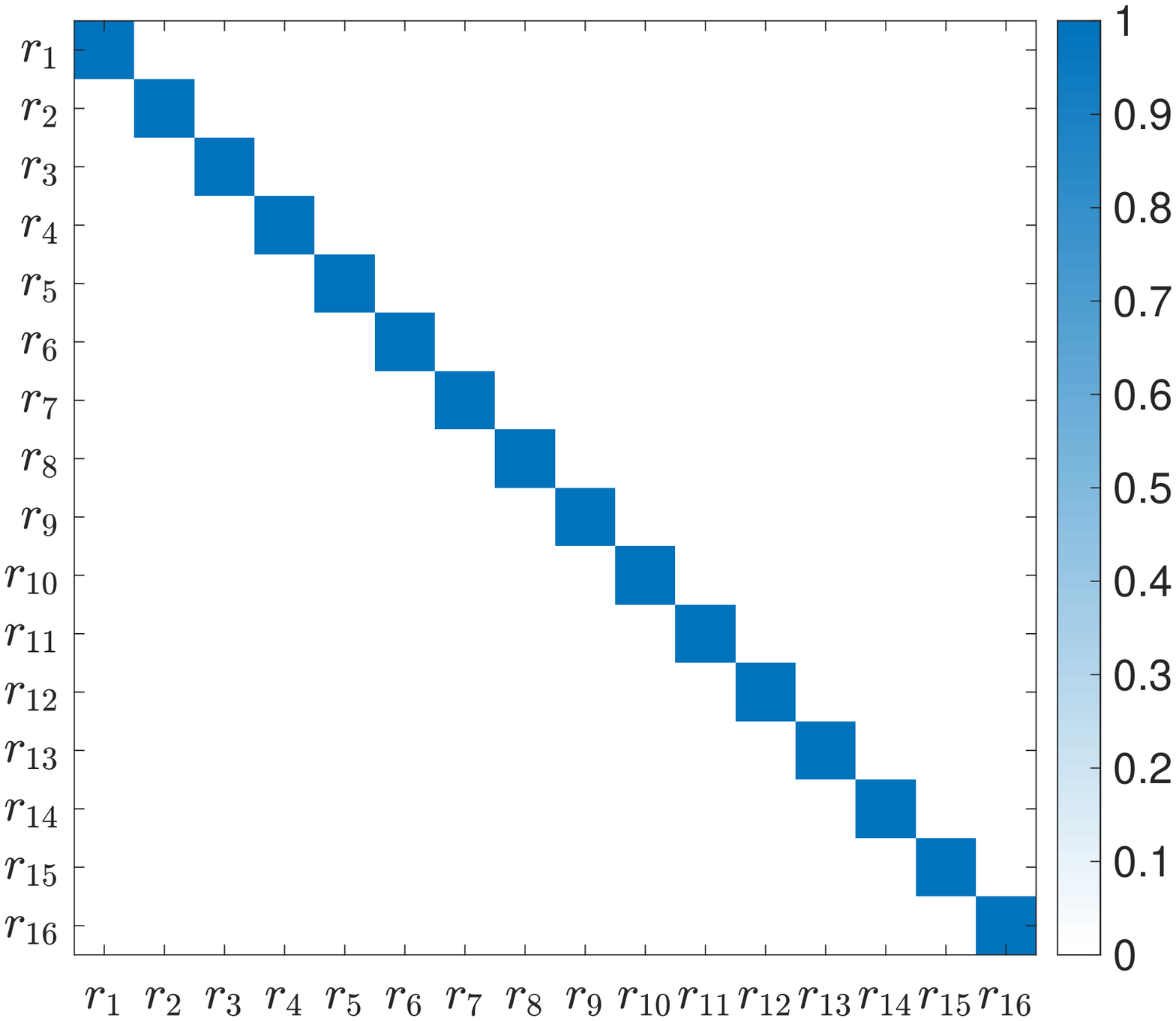}
        \caption{}
        \label{fig:4devices_Impairment_loc1}
    \end{subfigure}
     \begin{subfigure}{0.25\textwidth}
    \centering
        \includegraphics[width=\linewidth]{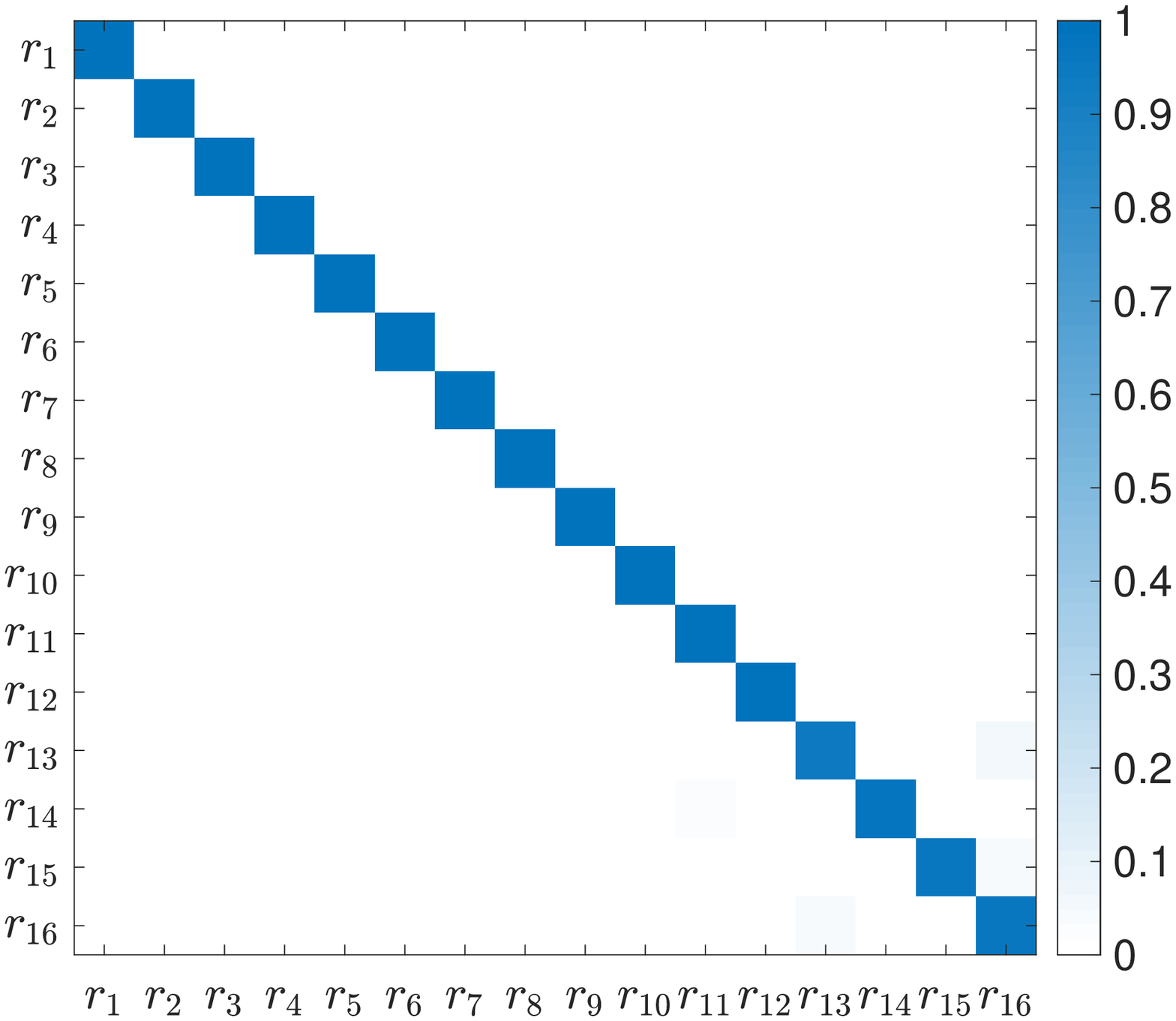}
        \caption{}
        \label{fig:4devices_Impairment_loc2}
    \end{subfigure}%
    \centering
    \begin{subfigure}{0.25\textwidth}
    \centering
        \includegraphics[width=\linewidth]{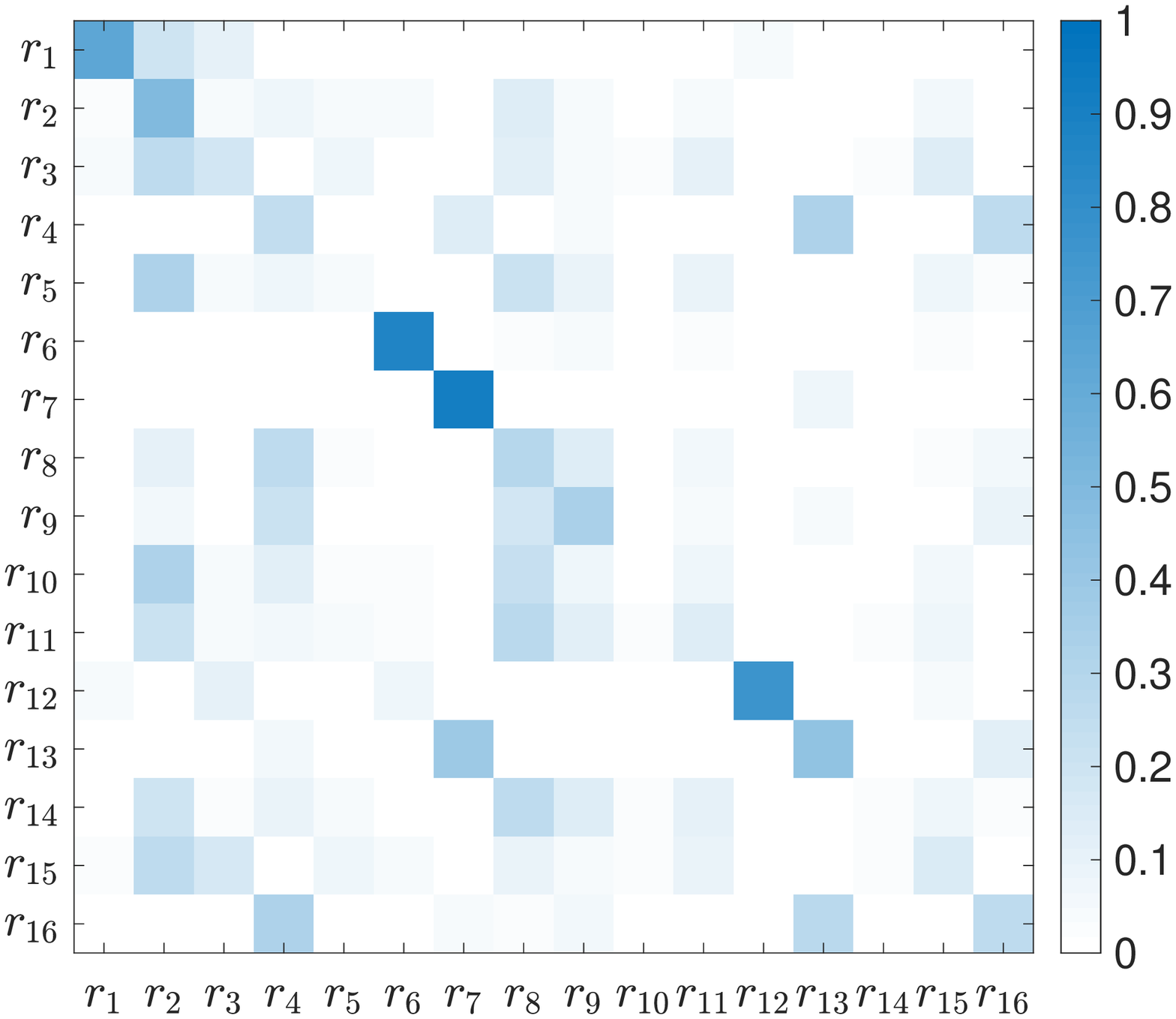}
        \caption{}
        \label{fig:4devices_NoImpairment}
    \end{subfigure}
    \caption{Classification accuracy (a) via cable; (b) over air in location 1 (Fig.\ref{fig:location1}); (c) over air in location 2 (Fig.\ref{fig:location2}). (d) shows the accuracy without ORACLE (data collected in location 2).}\vspace{-0.5cm}
    \end{figure}

We test the performance of the trained CNN classifier with $16$ X310 radios. To do so, we first collect samples from these radios through cable. All radios are configured with one of $16$ impairments selected from set $S$, according to the approach described in Sec.~\ref{sec:5d}.  As shown in Fig.~\ref{fig:4devices_Impairment}, ORACLE easily distinguishes bit-similar radios that are intentionally introduced with the selected impairments by achieving a classification accuracy of $99.76$\%. This indicates that our pre-trained classifier is able to identify bit-similar radios accurately.

Next, we evaluate the performance of ORACLE with data collected over the wireless channel. To show robustness to  variation in channel conditions, we conduct the experiments in two different locations: (1) our lab, which represents a typical in-indoor environment (Fig.~\ref{fig:location1}) and (2) a more open recreation area which has fewer reflections (Fig.~\ref{fig:location2}). 
The confusion matrix of classification accuracy is shown in Fig.~\ref{fig:4devices_Impairment_loc1} and Fig.~\ref{fig:4devices_Impairment_loc2} respectively. 
In general, in both environments ORACLE can achieve higher than $99.5$\% accuracy, which proves that the unique patterns created by the impairments can still be detected, even with random noise. 
 

In comparison, training the same classifier with these $16$ X310 devices without any kind of artificially introduced hardware impairments results in a poor classification performance.  As shown in Fig.~\ref{fig:4devices_NoImpairment}, the classification accuracy is only 35.96\% for these bit-similar radios, which shows the benefits of the careful impairment allocation process. 




\subsection{Reduced BER with heuristic impairments selection} \label{sec:6b}
\vspace{-4mm}
We use the metric of average total sum of BER of all the transmitters and compare the results with allocating impairments i) randomly, and ii) greedily using the algorithm described in Sec. \ref{sec:5d}. We consider  $R=4,8,12,16$ radios to have average SNR values selected randomly among \{20,25,30\} dB. Let IQ imbalance be the only impairment added, which is bounded by IMMR value of -13.5 dB. However, we consider 16 available impairment levels that range from IMMR of -13.5 to -21 dB with 0.5 dB separation. At each selection we ensure that the CNN classifies with these impairment levels at $>99\%$ accuracy. 

Under a random allocation approach, $R$ radios are randomly allocated one of the selected 16 impairment levels. On the other hand, our greedy heuristic algorithm iteratively assigns a lowest available impairment level to the radio which have least average  SNR level. 
A BER value for each radio is computed with different SNR levels shown in Fig.~\ref{fig:BERvsIQ}. We run 1000 iterations, in which each radio is randomly assigned one SNR level. In each iteration, a unique impairment level is randomly allocated to each radio using random allocation strategy. We repeat this 500 times to compute the total sum of BER of all the radios averaged over 500 iterations for the given SNR assignment. This is then averaged again over 1000 SNR assignments. Similarly, we compute the total sum of BER of all the radios obtained using the greedy heuristic algorithm, averaged over 1000 SNR assignments. Table \ref{Table:BERPerformance} shows the BER of all radios confirming that ORACLE's approach of allocating impairments always outperforms random allocation.  


\begin{table}[!h]
\caption{BER comparison between random and greedy heuristic impairments allocation.}
	\label{Table:BERPerformance}
\centering
\small
\begin{tabular}{|l|l|l|}
\hline
\multirow{2}{*}{Number of radios} & \multicolumn{2}{l|}{Average total sum of BER} \\ \cline{2-3} &  Random      &    Greedy Heuristic                 \\ \hline
$R = 4$  & $1.28 \times 10^{-3}$ &  $1.81 \times 10^{-5}$   \\ \hline
$R = 8$  & $2.62 \times 10^{-3}$ &  $7.82 \times 10^{-5}$   \\ \hline
$R = 12$ & $3.90 \times 10^{-3}$ &  $2.49 \times 10^{-4}$ \\ \hline
$R = 16$ & $5.20 \times 10^{-3}$ &  $8.13 \times 10^{-4}$ \\ \hline
\end{tabular}
\end{table}

%

\section{Conclusion}

We presented ORACLE, a fingerprinting technique for identification of specific radios based on the hardware-centric features within the transmitter chain. We showed that our CNN classier achieves an accuracy of $99\%$ using raw IQ samples for $>100+$ COTS WiFi devices and 16 X310 USRP radios in static environment. To further improve the classification accuracy in dynamic environment, we showed how feedback-driven transmitter-side modifications can increase differentiability for bit-similar devices. The key innovation lies in its `train once and deploy anywhere' feature. We demonstrate experimental $>99\%$ accuracy with bit-similar X310 radios, regardless of different channel conditions and wireless transmission environments. 

\section*{Acknowledgment}
This work is supported by DARPA under RFMLS program contract N00164-18-R-WQ80. We are grateful to Paul Tilghman, program manager at DARPA, and Esko Jaska for their insightful comments and suggestions.

\bibliographystyle{IEEEtran}    

\end{document}